\documentclass[12pt,preprint]{aastex}

\newcommand{\beq}	{\begin{equation}}
\newcommand{\eeq}	{\end{equation}}
\newcommand{\calm}	{{\cal M}}

\newcommand{\calr}	{{\cal R}}

\newcommand{\vecB}	{{\bf B}}

\newcommand{\veck}	{{\bf k}}
\newcommand{\vecv}	{{\bf v}}

\newcommand{\vecnabla}	{{\bf\nabla}}
\newcommand{\grad}      {{\bf\nabla}}
\newcommand{\gad}       {{\gamma_{\rm AD}}}
\newcommand{\gadp}      {{\gamma_{\rm AD,\, phys}}}

\newcommand{\curl}      {{\bf\nabla\times}}
\newcommand{\vectimes}	{{\bf \times}}
\newcommand{\alfven}    {{Alfv$\acute{\rm e}$n }}
\newcommand{\alfvenic}  {{Alfv$\acute{\rm e}$nic }}

\newcommand{\chio}	{\chi_{i0}}
\newcommand{\chiop}	{\chi_{i0,\rm\, phys}}
\newcommand{\mai}	{{\calm_{{\rm A}i}}}

\newcommand{\nh}	{n_{\rm H}}
\newcommand{\rad}	{R_{\rm AD}}
\newcommand{\meanrad}	{\langle\rad\rangle_V}
\newcommand{\meanradi}	{\langle\rad(\ell_{v_i})\rangle_V}

\newcommand{\tad}	{t_{\rm AD}}

\newcommand{\tni}	{t_{\rm ni}}
\newcommand{\tf}	{t_f}
\newcommand{\teq}	{t_{\rm eq}}
\newcommand{\ad}	{{\rm AD}}

\newcommand{\vai}	{v_{\rm Ai}}

\newcommand{\lvi}	{{\ell_{vi}}}

\newcommand{\avg}[1]    {{\langle #1 \rangle}}

\newcommand{\e}	        {$^{-1}$}

\newcommand{\eee}	{$^{-3}$}
\newcommand{\kphys}     {k_{\rm phys}}
\newcommand{\lad}        {{\ell_{\rm AD}}}

\newcommand{\ma}	{{\calm_{\rm A}}}

\newcommand{\nht}	{n_{\rm H,\, 3}}
\newcommand{\radl}	{R_{\rm AD}(\ell_0)}

\newcommand{\snt}       {\sigma_{\rm nt}}

\newcommand{\crion}	{\calr_{\rm ion}}
\newcommand{\crione}	{\calr_{\rm ion,\, eq}}

\shorttitle{MHD Turbulence Simulations with Ambipolar Diffusion}
\shortauthors{Li, McKee, \& Klein}
\begin{document}
\title{Sub-\alfvenic Non-Ideal MHD Turbulence Simulations with Ambipolar Diffusion: I. Turbulence Statistics}
\author{Pak Shing Li}
\affil{Astronomy Department, University of California,
    Berkeley, CA 94720}
\email{psli@astron.berkeley.edu}
\author{Christopher F. McKee}
\affil{Physics Department and Astronomy Department, University of California,
    Berkeley, CA 94720}
\email{cmckee@astro.berkeley.edu}
\author{Richard I. Klein}
\affil{Astronomy Department, University of California,
    Berkeley, CA 94720; and Lawrence Livermore National Laboratory,\\
    P.O.Box 808, L-23, Livermore, CA 94550}
\email{klein@astron.berkeley.edu}
\and
\author{Robert T. Fisher}
\affil{Flash Center, Department of Astronomy and Astrophysics, University of Chicago, 
    5640 S. Ellis Ave., Chicago, IL. 60637}
\email{rtfisher@uchicago.edu}

\begin{abstract}
Most numerical investigations on the role of magnetic fields in turbulent 
molecular clouds (MCs) are based on ideal magneto-hydrodynamics (MHD).  
However, MCs are weakly ionized, so that the time scale required for the 
magnetic field to diffuse through the neutral component of the plasma by 
ambipolar diffusion (AD) can be comparable to the dynamical time scale.  We 
have performed a series of $256^3$ and $512^3$ simulations on supersonic but 
sub-\alfvenic turbulent systems with AD using the Heavy-Ion Approximation developed in \citet{li06}.  
Our calculations are based on the assumption that the number of ions is conserved, but we show that these results approximately
apply to the case of time-dependent ionization in molecular clouds as well.
Convergence studies allow us to determine the optimal value of the ionization mass fraction when using the heavy-ion approximation for low Mach number, sub-\alfvenic turbulent 
systems.  We find that ambipolar diffusion steepens the velocity and magnetic 
power spectra compared to the ideal MHD case. Changes in the density PDF, total
 magnetic energy, and ionization fraction are determined as a function of the 
AD Reynolds number.  The power spectra for the neutral gas 
properties of a strongly magnetized medium with a low AD Reynolds number are 
similar to those for a weakly magnetized medium; in particular, the power 
spectrum of the neutral velocity is close to that for Burgers turbulence.
\end{abstract}
\section{Introduction}

Both supersonic turbulence and magnetic fields are widely
observed in molecular clouds (MCs).  MCs have broad line widths, ranging
from a few to more than 10 times the sound speed, $c_s$  \citep[]{elm04}.  The
observed interstellar magnetic field strength is a few micro
Gauss \citep[e.g.][]{hei05},  in rough equipartition with
the kinetic energy in the interstellar medium.  If the magnetic field were perfectly frozen
into the interstellar gas during the gravitational collapse of a protostar, the magnetic
field strength of a typical star like our Sun would be more than 10 orders
of magnitude larger than we observe today.  Thus  there must
be some mechanisms that are effective in removing the excess magnetic
flux during the star formation process.
\citet{mes56} first suggested that ambipolar diffusion (AD) could allow
magnetic flux to be redistributed during collapse in low
ionization regions in MCs as the result of the differential motion between the
ionized and neutral gas.  Since then, much work has been done on
AD-driven collapse of MCs
\citep[e.g.][]{spi68,nak72,mou76,mou77,mou79,nak78,shu83,liz89,fie92,fie93}.

AD-driven gravitational contraction is a quasi-static process.  The AD timescale, $\tad$,
in a typical MC is about ten times the free-fall time,
$t_{\rm ff}$ \citep{mck07}.  Once AD has removed a sufficient amount of magnetic flux,
a thermally supported core with a mass in excess of the Bonnor-Ebert mass will collapse in
about a free-fall time. However, MCs are observed to be supersonically turbulent.
Numerical simulations have shown that turbulence is an efficient  mechanism for supporting
MCs globally, while at the same time providing seeds for gravitational collapse
by shock compression.
\citep[e.g.][]{kle00,hei01,li04}.  By driving large fluctuations in 
the density, velocity and magnetic field, turbulence
significantly reduces the AD timescale \citep{fat02,zwe02}.  
AD significantly redistributes magnetic flux. While the importance of
this process has long been recognized in studies of star formation,
it is also important in the simpler case in which self-gravity is weak.
We therefore wish to determine how AD affects
the properties of supersonic, magnetized turbulence such as
that inferred in MCs.

It is very challenging to carry out three-dimensional (3D) simulations of AD in molecular clouds.
The small ionization fraction in molecular clouds means that the ion
inertia can be neglected. If the AD is treated as diffusion of the magnetic
field in a single fluid, the time step in explicit codes scales as the 
square of the grid-size ($\Delta x^2$), which is
prohibitive at high resolution; the time step is also
proportional to the ionization, making it impossible to
simulate the small ionizations found in MCs \citep{mac95}. 
Treating the ions and neutrals separately
as two fluids permits a time step proportional to $\Delta x$, but
the necessity of following the \alfven waves in the ion fluid again leads
to very small time steps. 
Two-dimensional fully-implicit
codes \citep[e.g.][]{fie92} were developed to avoid this problem, but
complex code development would be required to extend this to three dimensions.
In addition, implicit treatments can involve multiple iterations to
converge,  which may offset the advantage from the larger timestep.  Some
attempts have been made to perform 3D turbulence simulations with AD with
semi-implicit schemes \citep[e.g.][]{mac97,fal03} but with a heavy cost on
computational time.  

To overcome this problem, we introduced
the heavy-ion approximation \citep{li06}, in which the
ionization mass fraction is increased (so as to reduce
the ion \alfven velocity) and the ion-neutral collisional coupling
constant decreased, with the combined result that the ion-neutral drag
is unchanged.  With this
approximation, one can perform non-ideal MHD turbulence
simulations using a two-fluid approach while retaining an accurate 
treatment of the dynamical interaction between
the ions and neutrals in systems with realistic ionization fractions.
\citet{ois06} independently made this approximation
and used it to make a preliminary study of turbulence with AD.
Li et al. (2006) discussed the accuracy
of the heavy-ion approximation and developed criteria for the use of
this approximation in treating  MHD flows with AD.   

In this paper, we investigate the effects of AD on sub-\alfvenic turbulent
flows with a series of $256^3$ and $512^3$ MHD
turbulence simulations using ZEUS-MPAD (see Li et al. 2006) with the heavy-ion
approximation. The AD Reynolds number, $\rad$, is an important metric in the determination
of the significance of AD on turbulent flow (\citealp{zwe97}; Li et al. 2006)---for $\rad\gg 1$, the flow
is approximately an ideal MHD flow, whereas for $\rad\ll 1$ the flow of the neutrals
approaches a purely hydrodynamic flow. We wish to determine
the effect of varying the
AD Reynolds number on turbulence:  
How do the statistical properties of a
turbulent flow depend upon the AD Reynolds number as the flow changes from
the strong ideal MHD case to the strong AD case?   How do the effects of
turbulent driving of both the neutral and ion gas differ from driving of the neutral component alone?
What are the criteria necessary to achieve convergence in both the
spatial domain and in the simulation-time domain  when using the
heavy-ion approximation?  What value of the ionization fraction can be
used in this approximation and what errors are obtained as a function of
the ionization fraction used?  What is the effect of AD on the velocity
and the magnetic field power spectra?  How do these power spectra compare
with recent theoretical work on incompressible turbulence in a strong
magnetic field?  How do the power spectral indices compare with more
classical turbulent models for smooth, incompressible flows and shocked,
compressible flows?   It is well known that
the probability density function (PDF) for the density in supersonic
isothermal turbulence is log-normal.  Is this behavior valid in the presence of AD?  
Finally, how does the presence of ambipolar
diffusion in strongly magnetized turbulent clouds affect our
interpretation of the observed power spectra from these clouds?  

We discuss the heavy-ion approximation and the requirements for its validity in \S2.  
In \S3, we describe our models based on dimensionless model
parameters.  Because of the size of the parameter space of non-ideal MHD supersonic
turbulence, we focus our work on sub-\alfvenic turbulence with a thermal
Mach number $\calm$ = 3.  In \S4, we report our convergence study with the
heavy-ion approximation, investigating both spatial and temporal convergence.  In \S5, we report the results on the power
spectra of ion and neutral velocities and of the magnetic field in turbulence 
as a function of $\rad$.  In \S6, we discuss the probability density
function (PDF) of the gas density in the turbulent system.  In \S7, we
investigate other physical properties of the turbulent systems that scale
with $\rad$.  We summarize our results in \S8.
Most of the models investigated in details in this paper are based on the assumption
of ion conservation.  However, high density MCs generally have an ionization equilibrium timescale that is shorter than the dynamical timescale, so that the ionization inside MCs is most likely close to equilibrium.  In the Appendix, we demonstrate that nonetheless the assumption of ion conservation 
is generally a satisfactory approximation for molecular clouds.
The astrophysical implications of our turbulence simulations will be reported
in a subsequent paper.

\section{The Heavy-Ion Approximation}
In paper I, we formulated the heavy-ion approximation method for a two-fluid 
approach to MHD simulations with AD.  The isothermal MHD equations for the two fluids, ions and neutrals, with AD are:
\begin{eqnarray}
\frac{\partial \rho_n}{\partial t} &=& - \vecnabla \cdot (\rho_n
\vecv_n),
\label{eq:contn}\\
\frac{\partial \rho_i}{\partial t} &=& - \vecnabla \cdot (\rho_i \vecv_i),\\
\rho_n \frac{\partial \vecv_n}{\partial t} &=& - \rho_n 
(\vecv_n \cdot \vecnabla)
\vecv_n - \grad P_n - \gad \rho_i \rho_n (\vecv_n - \vecv_i),
\label{eq:nmotion}\\
\rho_i \frac{\partial \vecv_i}{\partial t} &=& - \rho_i (\vecv_i 
\cdot \vecnabla)\vecv_i 
- \grad P_i - \gad \rho_i \rho_n (\vecv_i - \vecv_n) 
+\frac{1}{4 \pi} \left( \curl \vecB \right) \vectimes \vecB,
\label{eq:imotion}\\
\frac{\partial \vecB}{\partial t} &=& \curl (\vecv_i \vectimes \vecB),
\label{eq:faraday}\\
\vecnabla \cdot \vecB &=& 0,
\label{eq:divb}
\end{eqnarray}
where $\rho$ = density, \vecv = velocity, \vecB = magnetic field strength, and $\gad$ = ion-neutral collisional coupling constant (note that in Paper I, this was denoted as $\gamma$).  
The subscripts $i$ and $n$ denote ions and neutrals, respectively.  Note that 
in this paper we do not include gravity, so the gravitational terms in the two momentum equations have been omitted.
In writing these equations, we have assumed that ions and neutrals are conserved, as in Paper I.  This is accurate only if the flow time is small compared to the recombination time. The inclusion of time-dependent chemistry brings in a number of uncertainties \citep{dal06}.
However, as we show in the Appendix,
the contribution from the ionization source terms is in general small compared to the AD drag term and can be ignored. 

We define the ionization mass fraction as
\beq
\chi_i\equiv\frac{\rho_i}{\rho_n},
\label{eq:chieq}
\eeq
which we assume to be small. 
For $\chi_i\ll 1$, as is the case in molecular
clouds, we could equally well define $\chi_i$ as the ratio
of $\rho_i$ to the total
density $\rho$, but in some of our numerical models we consider 
values of $\chi_i$ as large as 0.1, so that $\rho_n$ and $\rho$ are
not equivalent.  For simplicity, we set the physical value of the ionization mass fraction at a typical observed value of $\chiop=10^{-6}$ in our simulations.

In paper I, we followed \citet{mac97} in implementing a semi-implicit method for solving the momentum equations (\ref{eq:nmotion}) and (\ref{eq:imotion}) in the ZEUS-MP code.  The new code, ZEUS-MPAD, 
was tested with several standard AD problems in Paper I.  
For a two-fluid code,
the Courant condition restricts the 
timestep $\Delta t$ to be less than $\Delta x/\vai\propto \sqrt{\chi_i}$,
where $\vai$ is the ion \alfven velocity.  
For the low ionizations observed in molecular clouds, $\chi_i \sim 10^{-6}$, it is still not feasible to perform turbulence simulations even on the latest state-of-the-art supercomputing platforms.
This problem is compounded because
intermittency and the highly supersonic nature of the turbulence generate very large contrasts in density.  For the neutral component, the low density regions could have densities as low as $10^{-4} \rho_{n0}$ \citep[e.g.][]{li04}, and the ion density could be even lower. 
Therefore, we adopt the heavy-ion approximation developed in Paper I, in which the initial ion mass fraction $\chio$ is increased but the ion-neutral coupling coefficient $\gad$ is decreased so as to maintain the same product 
$\gad\chi_i=\gadp\chiop.$
Following the convention established in Paper I, we set the physical value of the ion-neutral coupling coefficient $\gadp=9.21\times 10^{13}$~cm$^3$~g\e~s\e,
so that our simulations have $\gad\chio
=9.21\times 10^7 $~cm$^3$~g\e~s\e.
The heavy-ion 
approximation reduces the frequency of \alfven waves in the ions, which correspondingly increases the integration timestep, but it maintains the same dynamical coupling between ions and neutrals.  We performed three tests
in Paper I---the formation of a C-shock, the Wardle instability, and a one-dimensional self-gravitating AD collapse---and demonstrated that MHD simulations with AD can be sped up by a factor of 10 to 100, depending on the problem, without seriously affecting the accuracy.
In this paper, we shall consider values of $\chio$ from $10^{-4}$ to $10^{-1}$---i.e.,
$10^2 - 10^5$ times greater than the typical physical value. We shall show that
$\chio=10^{-2}$, corresponding to a speed-up by a factor $\sim 100$, gives good accuracy.

The importance of AD to the flow on a length scale $\ell$ is determined by the ambipolar diffusion Reynolds number, 
\beq
R_{\rm AD}(\ell)\equiv\frac{\ell v}{\tni v_A^2}
=\frac{4\pi\gad\rho_i\rho_n \ell v}{\avg{B^2}}=\frac{\ell}{\ell_\ad};
\label{eq:rad}
\eeq
where $\ell_{\rm AD}$ is the AD length scale \citep{zwe97,zwe02}
The three tests in Paper I all had $\rad\sim 1$ on the length scale of the problem.
Supersonic turbulent flows have large contrasts in density, velocity, and magnetic field, and as a result there is a large range of length scales involved.  The length scales of the local magnetic and velocity fields are
\begin{eqnarray}
\ell_{\delta B} &=& \left|\frac{\delta B}{\grad \delta B}\right|,\\
\ell_{ v} &=& \left|\frac{ v}{\grad v}\right|,
\label{eq:lv}
\end{eqnarray}
where $\delta B$ is the change in magnetic field from the mean field $B$.
and where we have assumed that the mean velocity of the system is zero.
Since the ion inertia is negligible in the astrophysical problem, it is necessary to ensure that it remains small when the heavy ion approximation is used.
By comparing the inertia term and the AD drag term in the momentum equations, we deduced in Paper I that this requires
\beq
\rad(\ell_{v_i}) \gg \mai^2
\label{eq:valid}
\eeq
where $\ell_{v_i}\sim\ell_{\delta B}$.  We shall verify that this  condition is
satisfied in our simulations.

\section{Summary of Simulations}

In this paper, we present a series of scale-free turbulence simulations with AD.
Three dimensionless numbers characterize the simulations:
(1) $\calm\equiv 3^{1/2}\snt/c_s$, the 3D rms Mach number of the turbulence,
where $\snt$ is the 1D rms nonthermal velocity dispersion;
(2)   the plasma $\beta = 8\pi\rho c_s^2/\avg{B^2}$, which measures the
importance of the magnetic field; and
(3) $\radl$, the AD Reynolds number on the scale of the box,
which measures the importance of AD. The relative importance of
the magnetic field on the dynamics of the gas as a whole and on the dynamics of
the ions is described by the
\alfven Mach numbers,
\beq
\ma=(\beta/2)^{1/2}\calm,~~~~ \mai\simeq\chi_i^{1/2}\ma,
\eeq
where the expression for $\mai$ is based on the approximation $\rho_n\simeq\rho$;
note that we are defining $\beta$ and $\ma$ in terms of the rms magnetic field,
not the mean field.
In molecular clouds, $\ma$ is observed to be of order unity \citep{cru99},
whereas $\mai$ is much less than unity.

For $\radl\gg 1$, the dynamics on the scale of the box are described
by ideal MHD; if in addition, the \alfven Mach number is large ($\ma\gg 1$), then
the dynamics are approximately described by hydrodynamics. In the opposite
limit of weak coupling, $\radl\ll 1$, we expect the neutral component to be approximately
hydrodynamic, whereas the ions will approximate an ideal MHD fluid of their own.
Observe that insofar as the neutrals are concerned, the cases of low $\radl$ and low $\ma$
could be confused with the case of high $\radl$ and high $\ma$, since in both cases
the neutrals behave approximately hydrodynamically.

The principal goal of this paper is to trace the transition from ideal MHD to 
weak coupling in a turbulent medium by varying the AD Reynolds number of
the turbulent box. It is
not our intention to carry out a complete parameter survey, so we
have fixed the plasma-$\beta=0.1$ and have carried out most
of our runs with a thermal Mach number $\calm = 3$. The corresponding
\alfven Mach number is $\ma=0.67$, which is 
comparable to the observed values.
We have also carried out a few runs with
$\calm = 10$, corresponding to $\ma=2.2$.

The simulations are carried out in a cubic box of size $\ell_0$ in
each dimension. Periodic boundary conditions are applied in 
all three dimensions, with the intention of approximately representing a small portion of a
molecular cloud. The initial magnetic field is oriented along the $z$-axis.
We consider the case of turbulence driven 
according to the recipe described in \citet{mac99}: a Gaussian random velocity field with a flat power spectrum in the range $1 \leq k \leq 2$, where $k\equiv\ell_0/\lambda
=\kphys\ell_0/(2\pi)$; i.e., the spectrum extends over the
wavelength range $\ell_0\geq\lambda\geq\ell_0/2$.
Random phases and amplitudes are generated in a spherical shell in Fourier space and then transformed back into coordinate space to generate each component of the driving velocity perturbation.  When all three velocity components are obtained, the amplitude of the velocity is scaled to a desired initial root-mean-square (rms) velocity, $v_{\rm rms}$, which is defined by the chosen 3D rms Mach number, $\calm$, of the model.  
Both the ion and neutral components start with the same velocity field initially and are driven by a fixed driving pattern.  We carried out experiments using a variable driving pattern and found that the results are statistically indistinguishable from the fixed driving pattern results.  We also compared driving both the ion and neutral components with driving only the neutral component, and again found no statistically significant differences. Therefore, to simplify the study, we performed all the simulations using a fixed driving pattern applied to both the ions and the neutrals.

Table 1 lists the initial dimensionless parameters for the models we have calculated. 
The models are labeled as ``mxcy", where $x=3$ or 10 is the Mach number and
$y= |\log\,\chio|$ describes the ionization
adopted in the heavy-ion approximation.
For the $\calm=10$ models, $\rad(\ell_0/4)$ = 1, which is identical to the model 
used in \citet{ois06}.
We include several different values of the AD Reynolds number:
$\radl$, which is based on the box size and the mean Mach number;
$\avg{\rad(\ell_{v_n})}_V$, 
the volume average of $\rad$ based on
the neutral velocity; and $\avg{\rad(\ell_{v_i})}_V$, the volume average based on the ion velocity. 
For the Mach 3 models, the latter two agree to within about a factor 2,
whereas for the Mach 10 models the agreement is within about a factor 3.

\section{Convergence Study}
\label{convergence}

Simulations of driven turbulence must be converged in both spatial resolution,
as is the case in any hydrodynamic simulation, and
in total simulation time, which is needed to reach a steady state.
For AD simulations that use the heavy-ion approximation, we must also
ensure convergence in the mean ionization, $\chio$. As described in \S 3,
we adopt a mean physical value for the ionization mass fraction of
$\chiop=10^{-6}$, but in our simulations we use a larger value of $\chio$
and a smaller value of the ion-neutral coupling constant, $\gad$, such
that $\gad\chio=9.21\times 10^{7}$~cm$^3$~g\e~s\e\ is constant.

\begin{figure}
\epsscale{.80}
\plotone{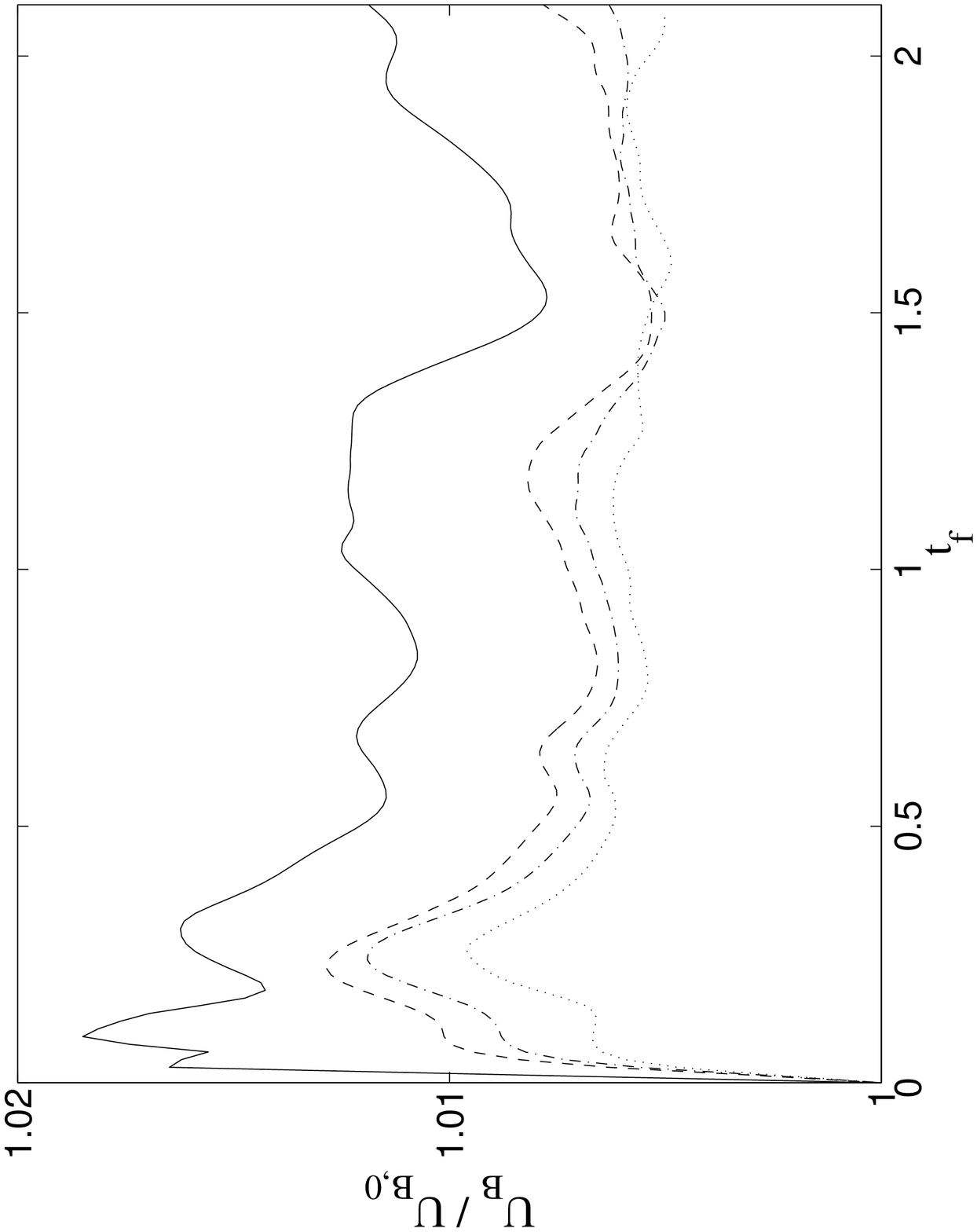}
\caption{Time evolution of the total magnetic energy, $U_B$, normalized to the initial total magnetic energy $U_{\rm B,0}$, for models m3c1 ($\chio = 10^{-1}$, solid line), m3c2 ($\chio = 10^{-2}$, dash line), m3c3 ($\chio = 10^{-3}$, dot-dash line), and m3c4 ($\chio = 10^{-4}$, dotted line).  The systems settle into approximate equilibrium states for $t\ga\tf$.
\label{fig1}}
\end{figure}

The convergence study performed in this section deals solely with globally-integrated quantities, such as the total magnetic energy.
Figure \ref{fig1} shows the results of a study of $\chio$-convergence, in which
we carried out runs with
$\chio= 10^{-1}$, 10$^{-2}$, 10$^{-3}$, and 10$^{-4}$ on a $256^{3}$ grid
with $\calm=3$. These four models show a similar evolution pattern, 
with an initial jump in the magnetic energy due to the initial perturbation
followed by evolution to a quasi-equilibrium with a fluctuating
magnetic energy.  Fluctuations in the magnetic energy in AD turbulence 
have been observed in other simulations as well \citep[e.g.][]{haw98}.  
These fluctuations appear to be random, and they prevent us from carrying out a
precise convergence study; in particular, we find that runs at different resolutions
or with different values of $\chio$ yield different time histories of the 
fluctuations.

\begin{figure}
\epsscale{.80}
\plotone{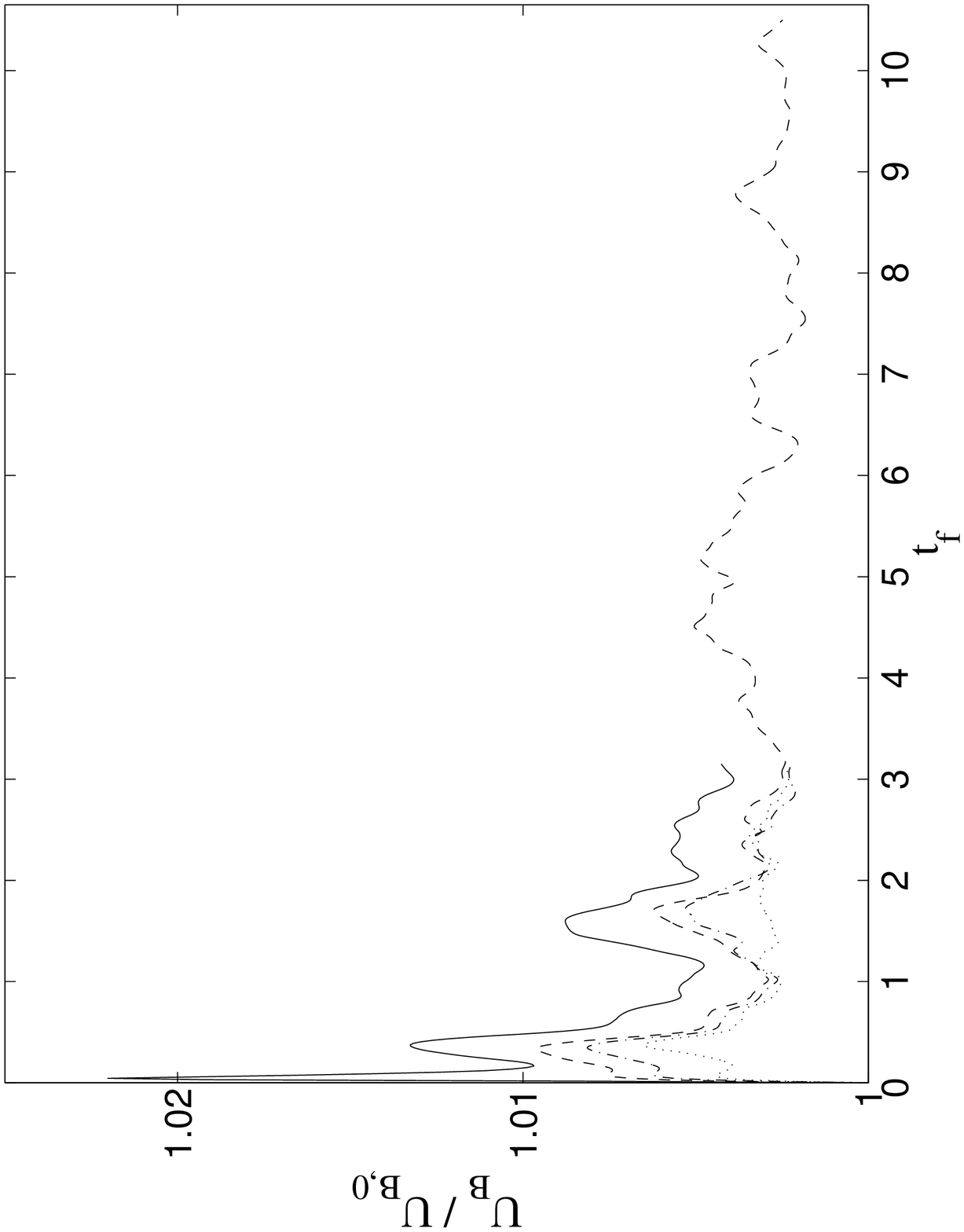}
\caption{Time evolution of the total magnetic energy, $U_B$, for four 128$^3$ turbulence models with the same initial conditions (models m3c1 to m3c4). One of the models (m3c2, with $\chio = 10^{-2}$) runs until $t>10\tf$.  The system is approximately in equilibrium for $t\ga\tf$, with random fluctuations.
\label{fig2}}
\end{figure}

\begin{figure}
\epsscale{.80}
\plotone{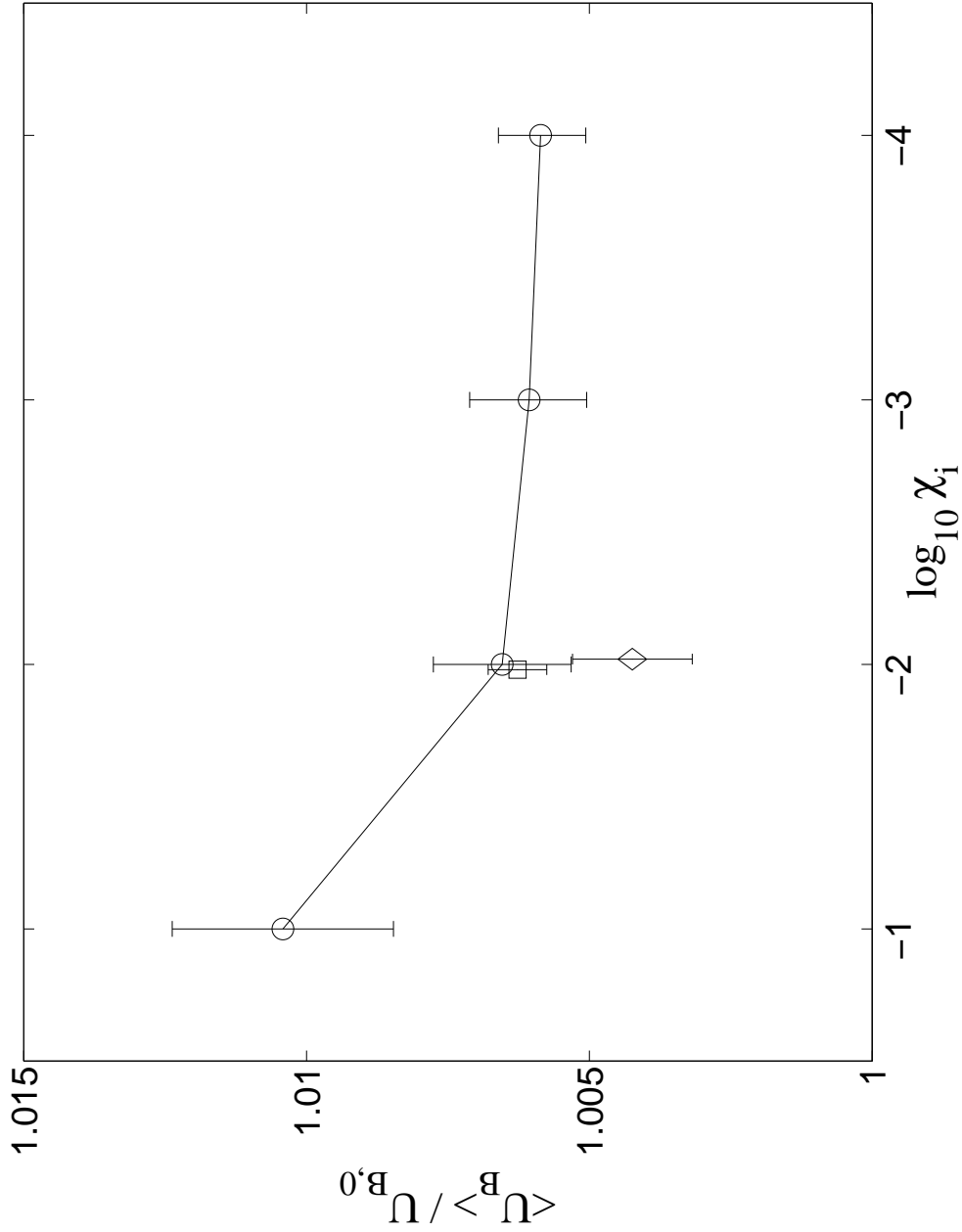}
\caption{Convergence behavior of the time-averaged total magnetic energy, $\langle U_B \rangle$, for models m3c1 to m3c4 as a function of $\chio$.  The total magnetic energies (circles) are averaged after the first crossing time and the error bars show the standard errors of the means.
The total magnetic energy is converged within the fluctuation limits for an ionization mass faction $\chio \leq 10^{-2}$.  The total magnetic energy of two models with the same initial conditions of the model m3c2 but with resolution of 128$^3$ (diamond) and 512$^3$ (square) are also plotted.  The $\chio$ of these two models are the same as m3c2 but changed here by a small amount in the plotting for the clarity of the overlapping error bars.
\label{fig3}}
\end{figure}

In order to address the issue of convergence in simulation time, we ran the model
m3c2 on a $128^3$ grid for a simulation time $>10 \tf$; such a long run is
prohibitive using a $256^3$ grid.  The flow time is defined as $\tf \equiv l_0/v_{\rm rms}$.
The time history of the magnetic energy is shown in Figure \ref{fig2}.  We see that the system is approximately in an equilibrium state after one flow time $\tf$, and that the fluctuations persist without a clear period.
Since the time to reach an equilibrium state is about $\tf$,
we continued the $\chio$-convergence simulations for a time
somewhat more than $2\tf$.  The mean and standard deviation of the magnetic energy determined after the first crossing time are shown in Figure \ref{fig3}.  We can see that the total magnetic energy converges quickly as $\chio$ decreases.  The variation of the mean magnetic energy among the models with $\chio = 10^{-2}$ to $10^{-4}$ is well within the amplitude of the fluctuations, but the error in the model with $\chio = 10^{-1}$ is larger than this. We conclude that the results are converged for $\chio\la 10^{-2}$.  In Paper I 
we showed that the
heavy-ion approximation is satisfied if the AD Reynolds number, $\rad$, is large compared to 
$\mai^2$ (eq. \ref{eq:valid}).  In a turbulent box simulation, it is not trivial to define $\mai$ or $\rad$ because of the intermittency of turbulence.   After driving the box for a period of time, the local $\mai$ and $\rad$ have enormous variations---for example, in the Mach 10 model with $\chio$ = $10^{-3}$, locally defined values of $\rad(\ell_{v_i})$ vary by a factor of 10$^{14}$!  Therefore, we compute the volume mean $\meanrad$ for both ions and neutrals, using the length
scale defined in equation (\ref{eq:lv}), and list them in Table 1.  The time at which these volume means are evaluated is also listed in the table.  Values of $\meanrad$ 
fluctuate, but vary by less than a factor of two for $t>t_f$ for 
both the Mach 3 and the Mach 10 models.  From Table 1, 
we see that the requirement to achieve a converged solution is actually 
$\mai^2/\meanradi\la 0.03$; this is well
satisfied for $\chio$ = 10$^{-2}$, which accounts for the accuracy of the
models with $\chio= 10^{-2}$ in Figure \ref{fig3}. Note that the
AD Reynolds number $\radl$ evaluated on the scale of the box is significantly larger than $\meanradi$, so having $\mai^2/\radl\ll1$ (or even $\la 0.03$)
is not sufficient for the validity of the heavy ion approximation.

To determine the spatial resolution required for turbulent AD simulations, we ran
models with 128$^3$ and 512$^3$ grid cells and with the same initial conditions as model m3c2.  The total magnetic energies of these two models are plotted in Figure \ref{fig3} alongside that of the 256$^3$ model.
The total magnetic energy, as well as other physical quantities, are converged at a resolution of 256$^3$ using ZEUS-MPAD.

From these convergence studies, we conclude that we can use the
heavy-ion approximation with $\chio = 10^{-2}$ to simulate systems with true 
values of $\chio\la 10^{-6}$.  A spatial resolution of at least $256^3$ is
needed. To obtain reliable statistical results, we suggest driving the system for more than 1
one flow time before measuring the physical quantities of the system.

\section{Power Spectra}
\label{sec:power}

The recent work of \citet{ois06} on MHD turbulence simulations with AD compared the magnetic energy spectra with and without ambipolar diffusion and concluded that AD produces no dissipation range in the magnetic energy spectrum.  In this section, we carry out a detailed investigation of velocity and magnetic field power spectra and show that AD does 
in fact have a small, but detectable, effect on the magnetic energy spectrum.

Generally, for an isotropic turbulent flow, the velocity power spectrum is computed using $P_v(k) = \Sigma \hat{u}_i(k)\hat{u}_i^*(k)$, where $\hat{u}_i(k)$ is the Fourier transform of the $i^{\rm th}$ component of velocity $u_i$(r) and the sum is over all three velocity components and all wave numbers $\veck$ in the 3D shell $k \leq |\veck| < k + dk$.  The inertial range of the power spectrum is expected to be a power law $P(k) \sim k^{-n}$.  For Kolmogorov \citep{kol41} and Burgers \citep{bur74} power spectra, $n = 5/3$ and 2, respectively.  Because of the relatively strong magnetic field for $\beta$ = 0.1, especially in our low Mach number models, highly anisotropic distributions are expected.  It is therefore necessary to compute both $P_v(k_r)$ and $P_v(k_z)$, the Fourier component power spectra perpendicular and parallel to the mean magnetic field, respectively, where $k_r = \sqrt{k_x^2+k_y^2}$.  For example, $P_v(k_r) = \Sigma \hat{u}_i(k_r)\hat{u}_i^*(k_r)$ where the sum is over all three velocity components and all wave numbers $\veck_r$ in the 2D shell $k_r \leq |\veck_r| < k_r + dk_r$ on the $x-y$ plane and over all planes along the cylinder with axis $z$.  The magnetic field power spectrum is calculated in the same manner.

Theoretical work on incompressible ideal MHD turbulence in a strong magnetic field \citep[e.g.][]{gol95,gol97,mar01} concludes that $P_v(k_\perp) \sim k_\perp^{-5/3}$, 
where $k_\perp$ is perpendicular to the {\it local} magnetic field;
this has the same exponent as the Kolmogorov spectrum.  
We express this in terms of the exponent in the power spectrum 
as $n_{vi}(k_\perp)\simeq 5/3$, where we have included the subscript
``$i$" to indicate that this applies to the ions.
However, numerical simulations \citep{mar01,mul03,bol05,ber06} give a flatter spectrum
that appears consistent with the Iroshnikov-Kraichnan 
spectrum, $n_{vi}(k_\perp)=3/2$ \citep{iro63,kra65}.
It is computationally much easier to calculate $P_v(k_r)$ than $P_v(k_\perp)$,
and fortunately the two are in close agreement
because $k_r \approx k_\perp[1+O(\theta^2)]$. On the other hand, $P_v(k_\parallel)$ 
is not similar to $P_v(k_z)$ because $k_z \approx \theta k_\perp$, where $\theta$ is the angle between the local magnetic field and the $z$-direction \citep{mar01},
so we shall not discuss $P_v(k_z)$.  Since the $k_z$ spectra are often not very meaningful, we also calculate the power spectra in terms of the total wavenumber, $P(k)$.  These spectra are also of interest because observers do not have information on the true direction of the magnetic field.  Any measurements of turbulent flows inside MCs will be restricted to the line of sight, which can be treated as a random direction from the true magnetic field direction.   We have demonstrated this by computing the power spectra $P_v(k_{60})$ and $P_B(k_{60})$ at 60$\degr$ from the $z$-axis (the median value of the angle relative to the field)
for the ideal MHD model m3i,  and we find that they agree with
the combined power spectra $P_v(k)$ and $P_B(k)$ to within the uncertainties.

\begin{figure}
\epsscale{.80}
\plotone{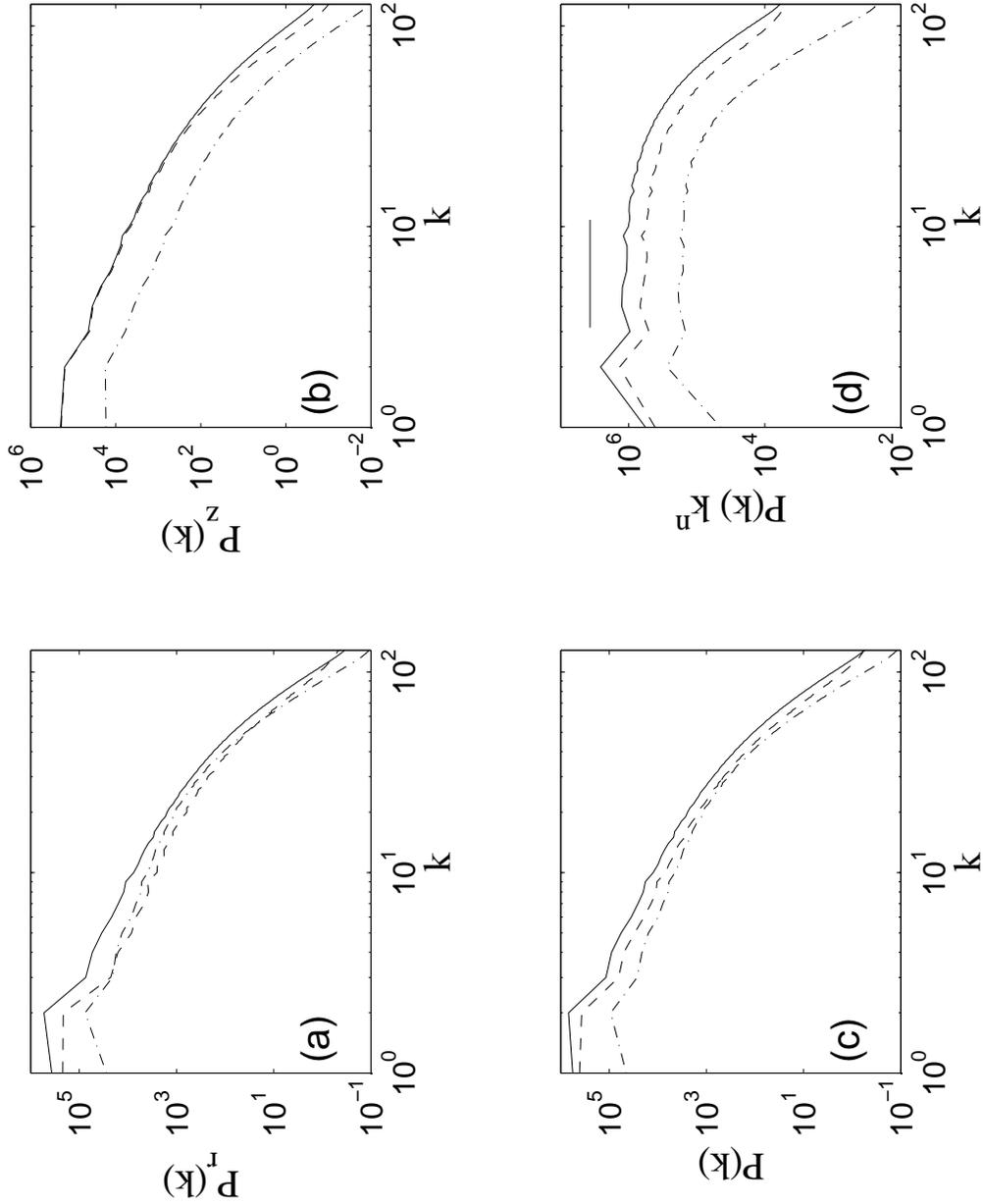}
\caption{(a) Velocity power spectra of the neutrals ($n_{vn,r}$, solid line) and ions ($n_{vi,r}$, dashed line) and the magnetic field power spectrum ($n_{B,r}$, dot-dash line), all perpendicular to the global magnetic field for model m3c2.  (b) Same as (a) but the components are parallel to the global magnetic field.  The ions and neutrals have very similar spectra parallel to the global magnetic field direction since only weak fields are induced in the perpendicular direction.  (c) The combined 3D velocity power spectra of neutrals ($n_{vn}$, solid line) and ions ($n_{vi}$, dashed line), and the power spectrum of the magnetic field ($n_B$, dot-dash line).  (d) The compensated 3D velocity power spectra of neutrals (solid line) and ions (dashed line), and the compensated magnetic field power spectrum (dot-dash line). The power law indexes used for the compensation are listed in Table 2.
\label{fig4}}
\end{figure}

In Figure \ref{fig4}, we present the time-averaged power spectra of the ion velocity and magnetic field for model m3c2 in the time interval $\sim (1-3)\tf$.  Because of the limited resolution (256$^3$), we do not expect the inertial range to extend much beyond $k$ = 10,
which corresponds to $\kphys \Delta x\simeq 0.25$.  Since the driving 
occurs between $k$ = 1 and 2, we have chosen to infer the power-law index by a least-squares fitting of the power spectrum from $k$ = 3 to 10.  
The uncertainty in the index is given by the standard error of the
mean, which we calculate as the standard deviation evaluated for a total 14 data sets between $1-3\tf$ divided by the square root of the number of independent samples of the index, which we estimate as 3.
In order to determine how long it takes for the turbulence system to become uncorrelated, we continued models m3i and m3c2 to a time somewhat greater than 5$\tf$.  
By studying the density correlation between data sets at different times, we found that they become essentially uncorrelated in a time slightly less than $\tf$.  We therefore take the number of independent samples to be the largest integer in $t_{\rm run}/t_f$.  For data sets dumped out between $1-3\tf$, we shall have 3 independent samples.  The range of wavenumbers used to determine the power-law
index of the power spectrum is very narrow, so
we carried out a high-resolution run with a resolution 
of $512^3$ (labeled m3c2h) and found that the power-law indexes agreed with those from the $256^3$ simulation within the errors.  We conclude that, 
although the results for the $256^3$ runs may not represent accurately the values
for the physical case in which the inertial range extends over many decades.
these results can be used to study the dependence of
the indexes on  the underlying physical parameters.

No time evolution of the power-law indexes is apparent for $t>t_f$: the time-averaged power indexes between $t_f$ and $2\tf$
agree with those between $2\tf$ and $3\tf$ within the uncertainties.  Figure \ref{fig4}a shows the power spectra, $P_{v,r}(k)$
and $P_{B,r}(k)$, of the of the velocity, $v_r$, and magnetic field, $B_r$, perpendicular to the global magnetic field.  Both ion and neutral velocity spectra are shown.  The close agreement between $P_{vi,r}(k)$ and $P_{B,r}(k)$ is consistent with equipartition between the ion kinetic and magnetic energy perpendicular to the mean field (\citealp{zwe95}; but see \S\ref{sec:scaling} below).  
Figure \ref{fig4}b shows the power spectra of the velocity and magnetic field parallel to the global magnetic field, $P_{v,z}(k)$ and $P_{B,z}(k)$.  
The power spectra for neutral and ion velocities parallel to the global field are about the same
because the weakness of the magnetic forces in this direction implies that the ions and neutrals are well coupled.  Figure \ref{fig4}c 
shows the combined power spectra $P(k)$ for the neutral and ion velocities and
for the magnetic field. The power-law indexes resulting from 
a least-squares fit to these spectra are listed in Table 2 and are used to produce
the compensated versions of these spectra in Figure \ref{fig4}d.

\begin{figure}
\epsscale{.80}
\plotone{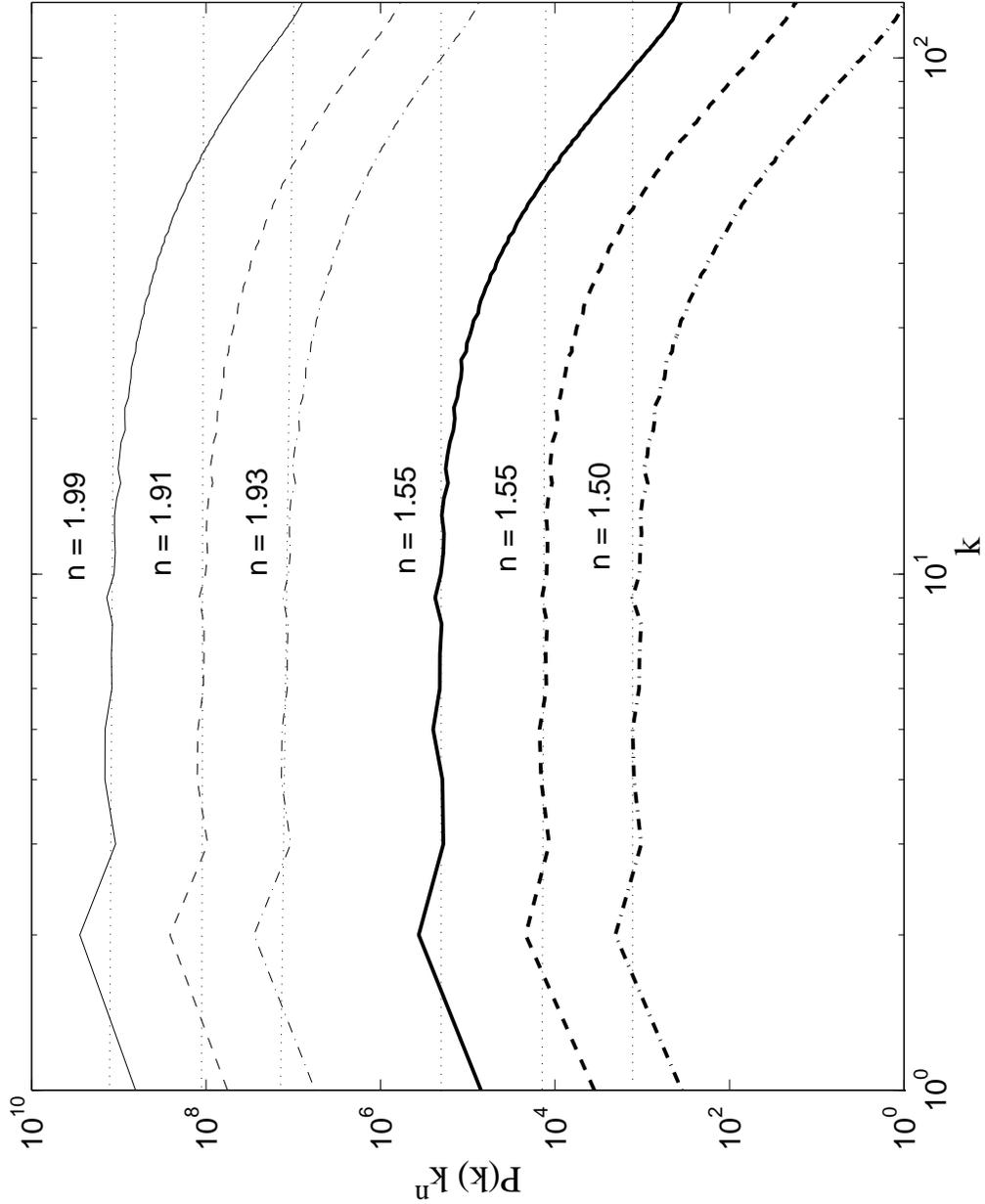}
\caption{The compensated power spectra of models m3c1 ($\chio = 10^{-1}$, solid line), m3c2 ($\chio = 10^{-2}$, dashed line), and m3c3 ($\chio = 10^{-3}$, dot-dash line). The spectra are compensated by the power law indexes of the inertial range fitted between $k = 3 - 10$.  The thin lines are neutral velocity spectra and the thick lines are the magnetic field power spectra.  The spectra are shifted up and down for side-by-side comparison.  As shown in the figure and listed in Table 2, the power law indexes are not sensitive to the choice of ionization mass fraction $\chio$, even for $\chio$ as large as 0.1.
\label{fig5}}
\end{figure}

First, we look at the $\chio$-convergence of the power spectra, using the heavy-ion approximation.  
As shown in Table 2, the power-law indexes of the four models m3c1 to m3c4,
which have $\chio=10^{-1}-10^{-4}$, are similar within the uncertainties. 
Figure \ref{fig5} shows 
this result graphically. Interestingly, the correct value of the power-law index can be obtained with $\chio=0.1$ even though we found in \S \ref{convergence} that convergence in magnetic field energy required $\chio\la 0.01$.
This lack of sensitivity to the value of $\chio$ could be 
because the relatively low resolution of the simulations and the intrinsic fluctuations discussed above lead to significant uncertainties in determining the power-law indexes. 

Next, we investigate whether driving the neutrals alone would alter the power spectrum of the ion (model m3c2a). In this case, the motion of the ions is mainly due to the drag force exerted by the neutrals. The spectral indexes are basically the same as in model m3c2, in which both the
ions and the neutrals are driven (see Table 2).  We conclude that our results are
insensitive to whether the driving applies to both the neutrals and ions or to the neutrals
alone.

How do the spectral indexes depend on the AD Reynolds number?
\citet{ois06} addressed this question by comparing a run with
$\rad(\ell_0/4) = 1$ to an ideal MHD run, both at Mach 10.
From visual inspection of their results, they did not find any significant difference
in the magnetic power spectra [$n_B(k)$---see their figure 3].  By contrast, when they compared a simulation with ohmic dissipation to an ideal MHD simulation, 
they found a large difference in the power spectra.
We performed three Mach 10 simulations with AD (models m10c1 to m10c3) using the same initial conditions as in \citet{ois06} and a Mach 10 ideal MHD model m10i for comparison.  The ionization mass fraction, $\chi_i$, in model m10c1 is 0.1, 
which is the same as in the AD model of \citet{ois06}.   Unfortunately, due to the low densities created in highly supersonic 
turbulence, it is computationally too expensive to continue the Mach 10 models much
beyond $t=\tf$.  Therefore, we obtained only one snapshot
of the turbulence for each case of the Mach 10 turbulence with AD, which is insufficient
to determine the uncertainty in the indexes. 

We then carried out a quantitative comparison between the AD
models and an ideal model at Mach 3. The magnetic spectral index for Model
m3i, an ideal MHD model with the same initial condition as the AD model m3c2,
is $n_B(k)=1.25\pm$0.09, which is clearly flatter than the value 1.55$\pm$0.12 
for the AD model.  We confirmed this result by comparing a high-resolution ideal
MHD model (m3ih at 512$^3$ resolution) with the high-resolution AD model
m3c2h.  However, the effect due to AD is much smaller than that of ohmic diffusion as reported in \citet{ois06}.

More generally, as shown in Table 3, the spectral indexes change systematically with the
AD Reynolds number. All the indexes listed undergo a statistically
significant increase in going from the ideal MHD case to the strongest AD case
[$\radl=0.12$].
The indexes for the neutral velocity increase down to the lowest value
of $\rad$, becoming slightly greater than 2; presumably they level
off at yet lower values of $\rad$, since they approach 
Burgers value of 2 in the hydrodynamic limit \citep{pad07}.
The index for the $z$-component of the ion velocity [$n_{vi,z}(k)$] is locked
to that of the neutral velocity since the ions and neutrals are well-coupled
parallel to the field. With the exception of $n_{vi,z}$, all the ion and magnetic
field indexes approach constant values at low $\rad$, although the value
of $\rad$ at which they level off varies. At the lowest value of $\rad$,
the index for the magnetic field has the Iroshnikov-Kraichnan value,
$n_B(k)=1.50 \pm 0.10$, 
as expected for strong-field (i.e., low-$\ma$) turbulence \citep{mar01,mul03,bol05,ber06}.
However, the effect of the neutrals on the ions
is still apparent in this strong AD case, since the ion velocity indexes differ significantly from
the ideal MHD values. This is to be expected, since
in the ideal MHD simulation, the power spectrum in the ion
fluctuations at small scales is due entirely to a cascade from larger scales, whereas
in the AD simulations the ions are driven at all scales by interactions with the dominant neutrals.  Therefore, for $\radl\la 1$, the power spectra of the neutral velocities and of at least the $z$-component of the ion velocity is close to a Burgers spectrum, and the power spectrum of the magnetic field is 
close to the Iroshnikov-Kraichnan spectrum.

\citet{pas88} pointed out that the observed linewidth-size scaling, $\sigma_v \propto \ell^{1/2}$ \citep[e.g.][]{sol87}, is what would be predicted for Burgers turbulence.  \citet{pad00} explained this as the result of ideal MHD turbulence in a weakly magnetized, supersonic (and hence super-Alfv$\acute{\rm e}$nic) medium.  Our results show that the Burgers spectrum can be obtained even in sub-\alfvenic turbulence if the AD effect is strong.
However, it must be borne in mind that our results apply to the inertial range in 
$256^3$ simulations, and as discussed above they differ by an unknown amount
from the physical case with a far larger inertial range.
Furthermore, the run with the lowest AD Reynolds number,
$\rad(\lvi)=0.015$, has a value of $\mai^2/\rad(\lvi)$
comparable to that for the $\chio=0.1$ runs, which are known to be
not fully converged; hence, we cannot be sure that this run is fully converged either.
The trends should be reliable, however.

\section{Probability Density Function}
\label{sec:pdf}

The probability density function (PDF) for the density of supersonic, isothermal
turbulence is log-normal \citep{vaz94}. That is, the volume-weighted or mass-weighted probability 
that the density has a given value is 
\beq
f_{V,\,M}\propto\exp[-(x\pm\mu_x)^2/2\sigma_x^2],
\eeq
where $x\equiv\ln(\rho/\bar\rho)$,  $\mu_x=\sigma_x^2/2$, and the plus and minus  
signs refer to the volume-weighted and mass-weighed probabilities, respectively
(e.g., \citealp{mck07}). Hence, the standard deviation of the distribution, $\sigma_x$,
is related to the means by
\beq
\avg{x}_M=-\avg{x}_V=\frac 12 \sigma_x^2.
\eeq

Table 4 gives the values for these quantities and for the median of $-x$ 
(labeled $-\tilde{x}$) for a range of values of $\radl$, extending from ideal MHD
[$\radl\rightarrow\infty$] to almost decoupled [$\radl=0.12$].
All these quantities should be equal for a log-normal PDF, and indeed they agree to
within the uncertainties, demonstrating that the log-normal behavior of the PDF is preserved in the case of ambipolar diffusion. (Note that the equality of the median and the mean confirms only that the PDF is symmetric, not that it is a log-normal.)  

\begin{figure}
\epsscale{.60}
\plotone{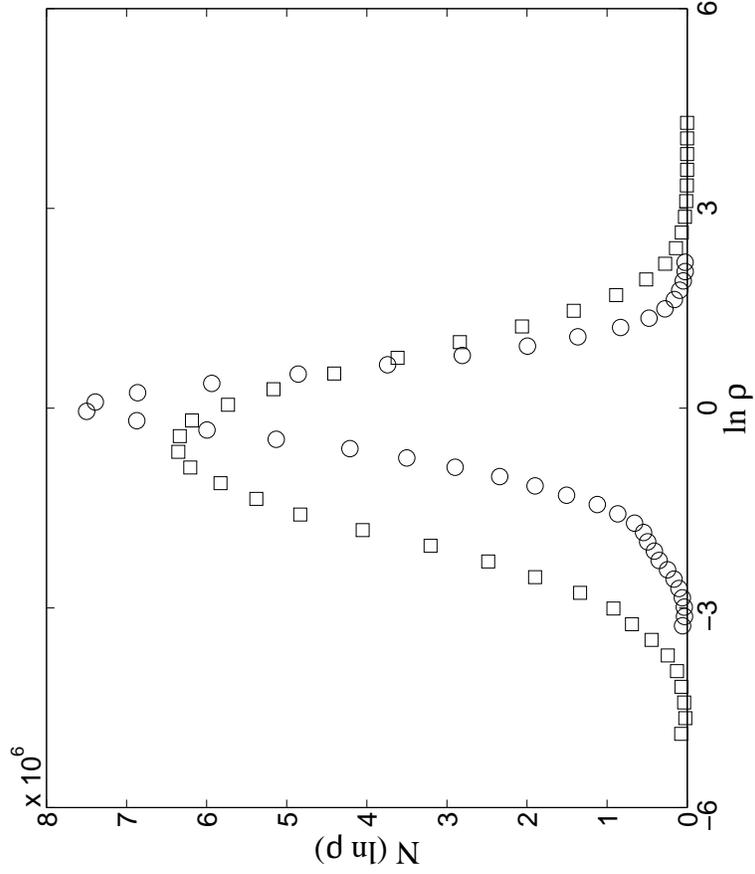}
\caption{Density PDF of ideal MHD model m3i (circles) and model m3c2 (squares).  The density PDF of model m3c2 shows significantly larger dispersion than the ideal MHD model because of the effects of ambipolar diffusion.  The dispersion and mean of the PDFs are listed in Table 4.
\label{fig6}}
\end{figure}

Table 4 shows, and Figure \ref{fig6} confirms, that the width of the density PDF increases
as AD becomes more important, although our results do not show a monotonic
behavior.
This increase is plausible due to the decreasing ability of the magnetic field to
cushion the shocks as $\radl$ decreases. For very small $\radl$, we expect the
dispersion to approach the hydrodynamic value,
\beq
\sigma_x^2 = {\rm ln}(1+\frac 14 \calm^2)
\eeq
\citep{pad02}. This corresponds to $\sigma_x^2/2=0.59$ for $\calm=3$, which is consistent with our
run at the lowest value of $\radl$. 
For the ideal MHD case, \citet{pad07} find a similar relation with the
sonic Mach number replaced by the \alfven Mach number.  \citet{ost01} do not find such a relation for this case, nor do we. It must be borne in mind that the simulations of Ostriker et al. (2001) had a range of values of $\beta$, and that our simulations all have $\beta=0.1$, which is substantially smaller than the value in the super-\alfvenic simulations of \citet{pad07}; in addition, our resolution is substantially less.

\section{Scaling with $\rad$}
\label{sec:scaling}

As we have seen in \S\ref{sec:power} and \S\ref{sec:pdf}, the statistical properties of the turbulent box vary as the AD Reynolds number of the system, $\radl$, changes from the ideal MHD case to the strong AD cases.  Here we shall examine several other properties as functions of $\radl$.

\begin{figure}
\epsscale{.80}
\plotone{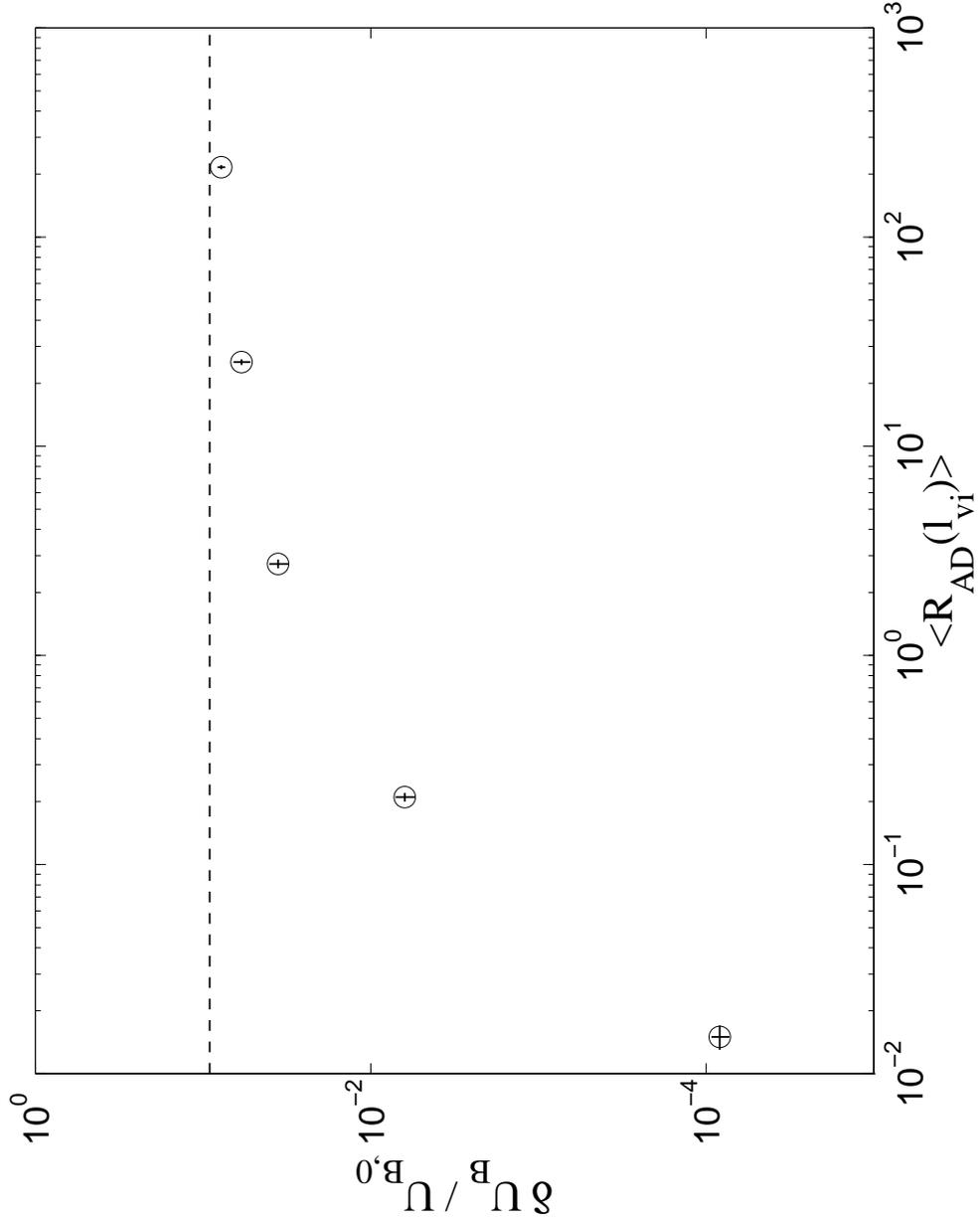}
\caption{Time-averaged change in the normalized total magnetic energy, $\delta U_B/U_{B0}$ for five models m3c2r-1 to m3c2r3 as a function of $\rad(\ell_{v_i})$.  Uncertainties are shown as error bars.  The dashed line is $\delta U_B/U_{B0}$ for the ideal MHD model m3i.  With increasing $\rad$, $\delta U_B/U_{B0}$ approaches the ideal MHD model value.
\label{fig7}}
\end{figure}

In the convergence study in \S\ref{convergence}, we used the total magnetic energy of the system to gauge the convergence in terms of spatial resolution and of ionization mass fraction, 
$\chio$.  Using the initial magnetic energy of the system as a reference, we can see how the fluctuating magnetic energy of the system, $\delta U_B = U_B - U_{B_0}$, changes as a function of $\rad$.  Figure \ref{fig7} shows that when $\rad$ is small, $\delta U_B$ is also small; indeed,
when $\rad$ approaches zero, that is when the ion fluid is totally decoupled from the neutral fluid, $\delta U_B$ should approach the value appropriate for the driven ions (recall that we drive both ions and neutrals).  Were we to drive only the neutrals, $\delta U_B$ would approach zero as $\rad\rightarrow 0$.

On the other hand, when $\rad$ increases, $\delta U_B$ increases to the value appropriate for an ideal MHD system. 
Since MHD waves have equipartition between the kinetic energy normal
to the field, $(1/2)\rho v_\perp^2$, and the perturbed magnetic energy, $\delta U_B$,
we expect $\delta U_B/U_{B0}=(2/3)\ma^2$, 
under the assumption that the
velocities are isotropic. Indeed, in their low-$\beta$ runs, \cite{sto98}
found $\delta U_B/U_{B0}\simeq 0.6\ma^2$, consistent with this expectation.
For our models, $\ma^2=\beta\calm^2/2=9/20$, so the theoretical expectation
is $\delta U_B/U_{B0}=0.3$. 
We find a significantly smaller value, however, $\delta U_B/U_{B0}\simeq 0.1$.
We attribute this to our boundary conditions: it is possible to have significant kinetic
energy that does not perturb the field, in the form of eddies rotating around the
field lines or flows along field lines. This effect was much smaller for \citet{sto98}
since they had a much smaller driving scale, peaked at $k=8$.
To see whether this effect could be significant in our models, we evaluated
$\avg{\vecv}^2$, where the average is over time and the results are
summed over all cells in the box. One would expect this to be close
to zero, whereas we found it to be a significant fraction of  $\avg{v^2}$.
Since these motions are at $k\sim 1$, however, they do not affect the power spectra.

\begin{figure}
\epsscale{.80}
\plotone{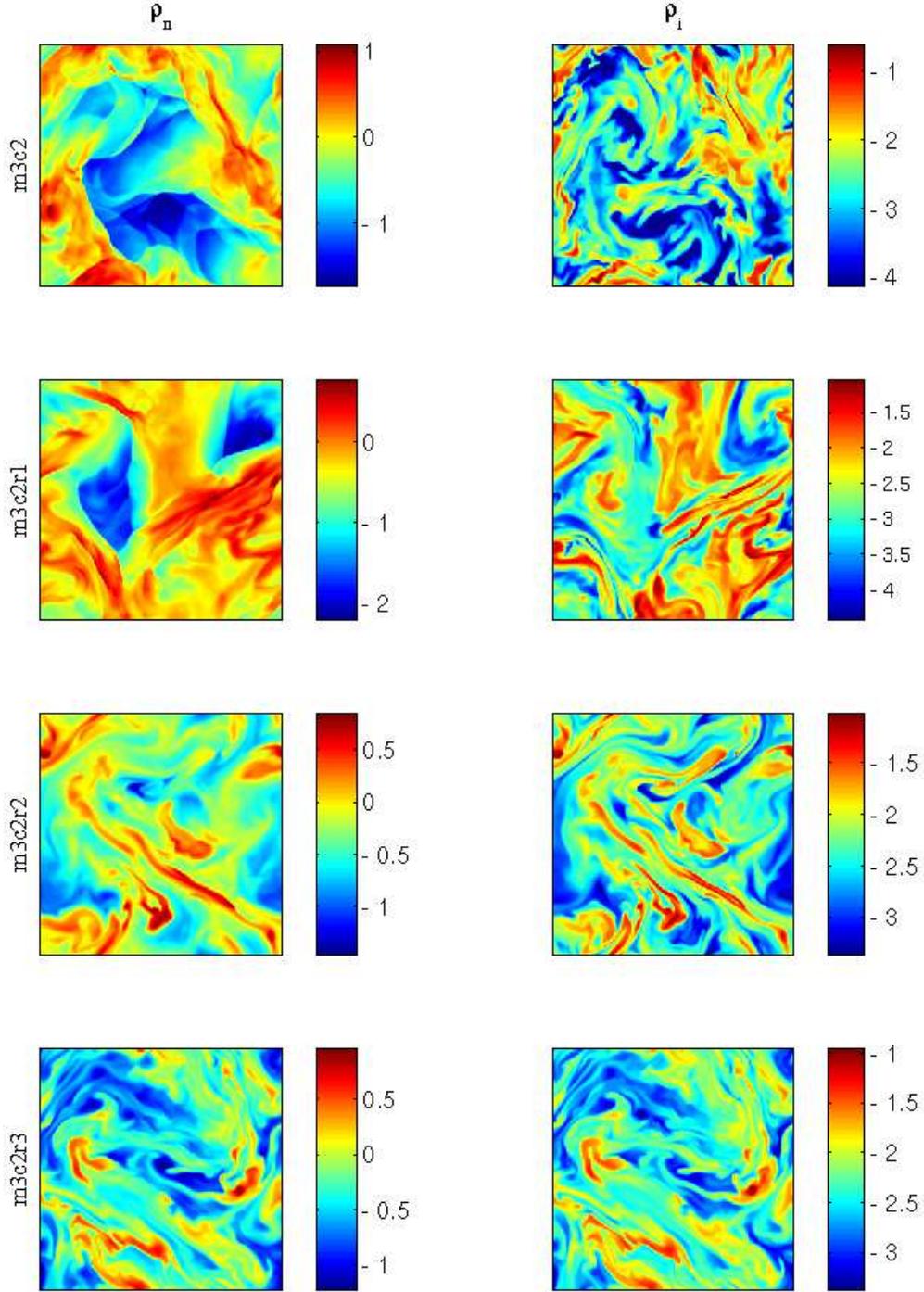}
\caption{Logarithmic density (log$\rho$) slices of models m3c2 (1st row), m3c2r1 (2nd row), m3c2r2 (3rd row), and m3c2r3 (4th row) at the middle of the turbulent box normal to the $z$-direction at time $t = 3\tf$.  The left column shows the neutral density and the right column shows the ion density.  When $\meanrad$ is large, the ions and neutrals are sufficiently strongly coupled that they evolve like a single fluid.
\label{fig8}}
\end{figure}

\begin{figure}
\epsscale{.80}
\plotone{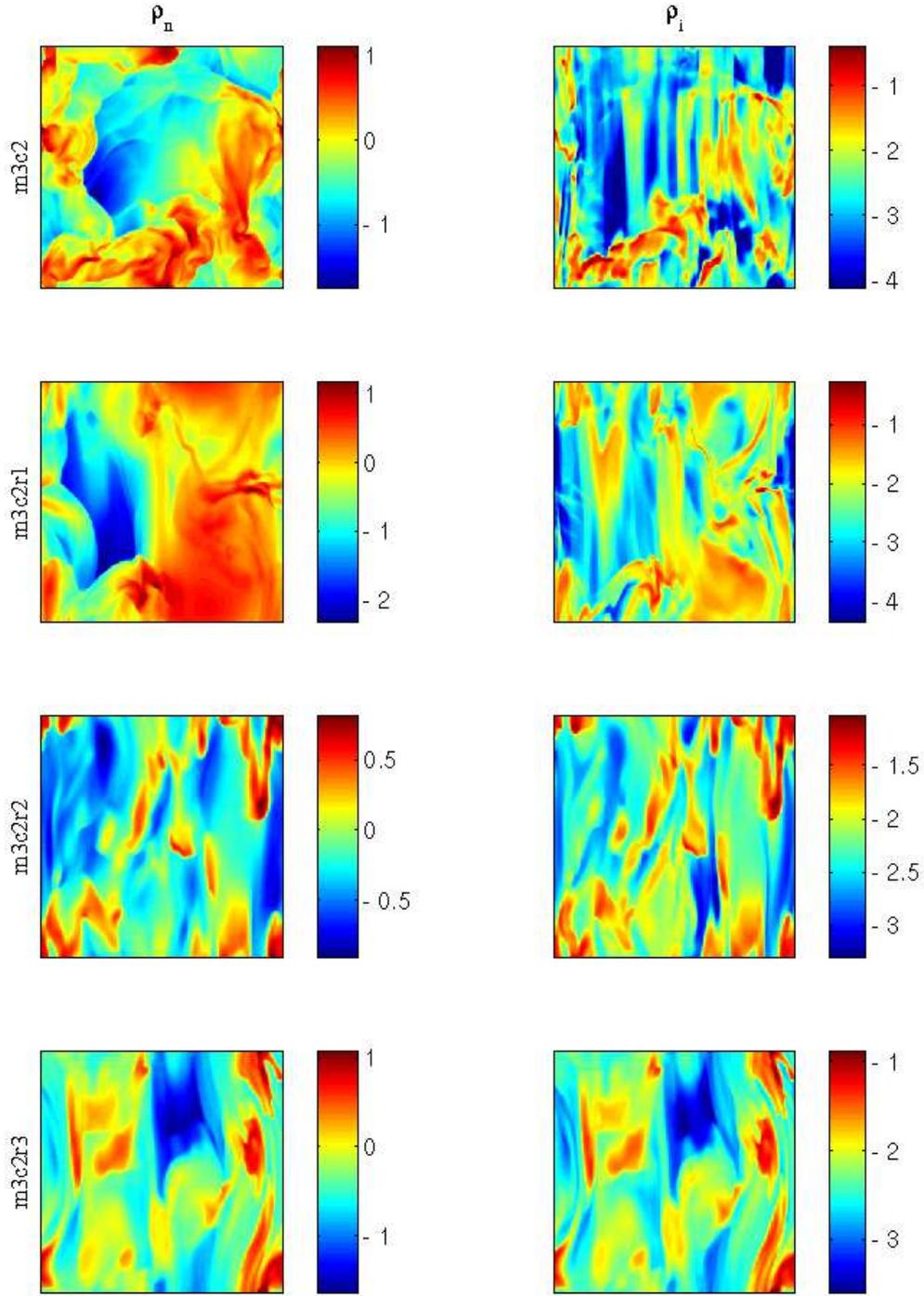}
\caption{Same as Figure \ref{fig8} but the slices are at the middle of the turbulent box normal to the $y$-direction.  The ion density is highly anisotropic due to the strong magnetic field and relatively weak turbulence ($\ma <1$).
\label{fig9}}
\end{figure}

Figures \ref{fig8} and \ref{fig9} show density slices for the ions and neutrals normal to the $z$ and $y$ axes, respectively, for models m3c2, m3c2r1, m3c2r2, and m3cr3 at $t = 3\tf$.  We see that the coupling between ions and neutrals gets stronger with larger $\rad$.  In Model m3c2r3, which has $\radl=1200$ and is very close to ideal MHD, the coupling between the ions and neutrals is so strong that there is hardly any difference between the spatial distributions of their densities.

\begin{figure}
\epsscale{.50}
\plotone{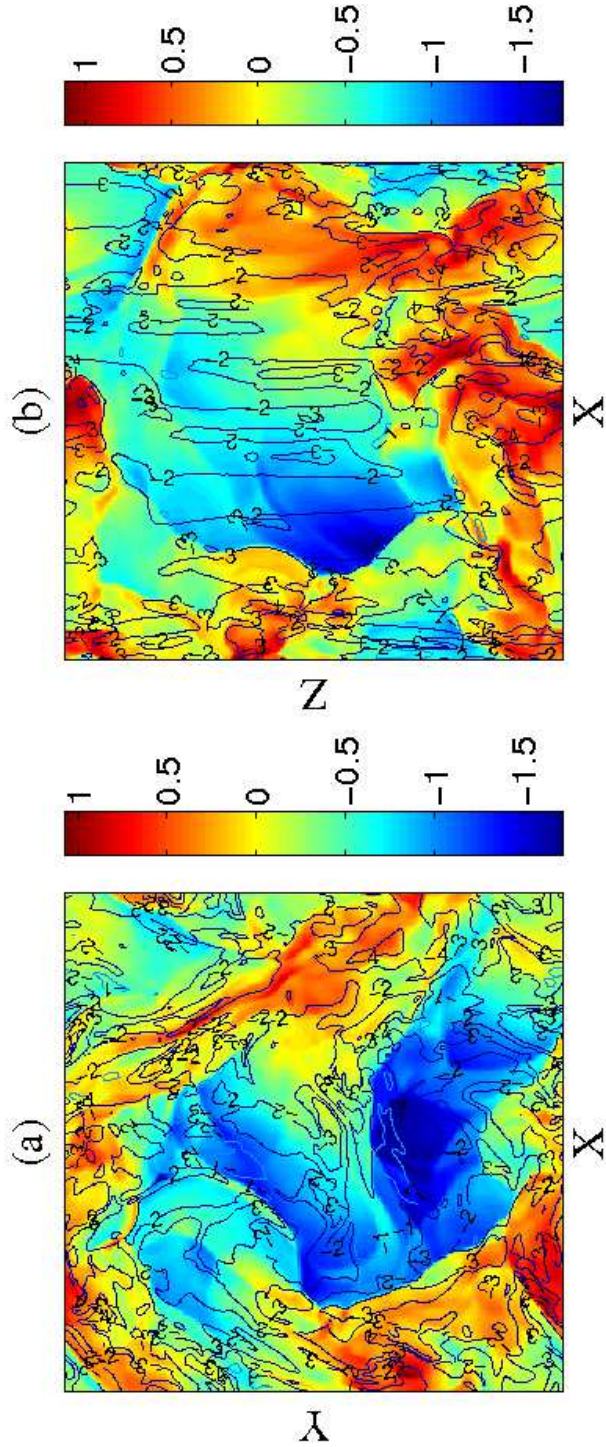}
\caption{Spatial distribution of ionization mass fraction $\chi_i$.  (a) The $\chi_i$ of a slice at the middle of the turbulence box normal to the z-direction (B-field direction).  The contours are log $\chi_i$ and the grey scale (color scale in online version) map is log $\rho_n$.  Small $\chi_i$ regions usally associate with high density regions.  (b) Same as (a) but the slice is normal to the y-direction.  The contours are highly anisotropic because of the restraint of ions due to strong magnetic field.
\label{fig10}}
\end{figure}

As mentioned above, supersonic turbulence quickly creates large density contrasts in the system.  In the presence of AD, it also creates large ionization contrasts, as can be inferred from Figures \ref{fig8} and \ref{fig9}.  
Figures \ref{fig10}a and \ref{fig10}b show the ionization mass fraction $\chi_i$ of a slice at the middle of the turbulence box normal to the $z$ and $y$
axes, respectively, from the model m3c2 at time $t = 3\tf$.  Regions of low ionization are found to occur in regions of high density: Ambipolar diffusion allows shocks to compress the neutrals much more than the ions. Because we have assumed overall ion conservation, the mass of ions on a given flux tube is constant in the absence of numerical diffusion. Therefore, the change in $\chi_i$ is purely a dynamical result. Furthermore, the contours of $\log\rho_i$ are highly anisotropic and are aligned with the $z$-axis, as shown in Figure \ref{fig10}b.  

\begin{figure}
\epsscale{.50}
\plotone{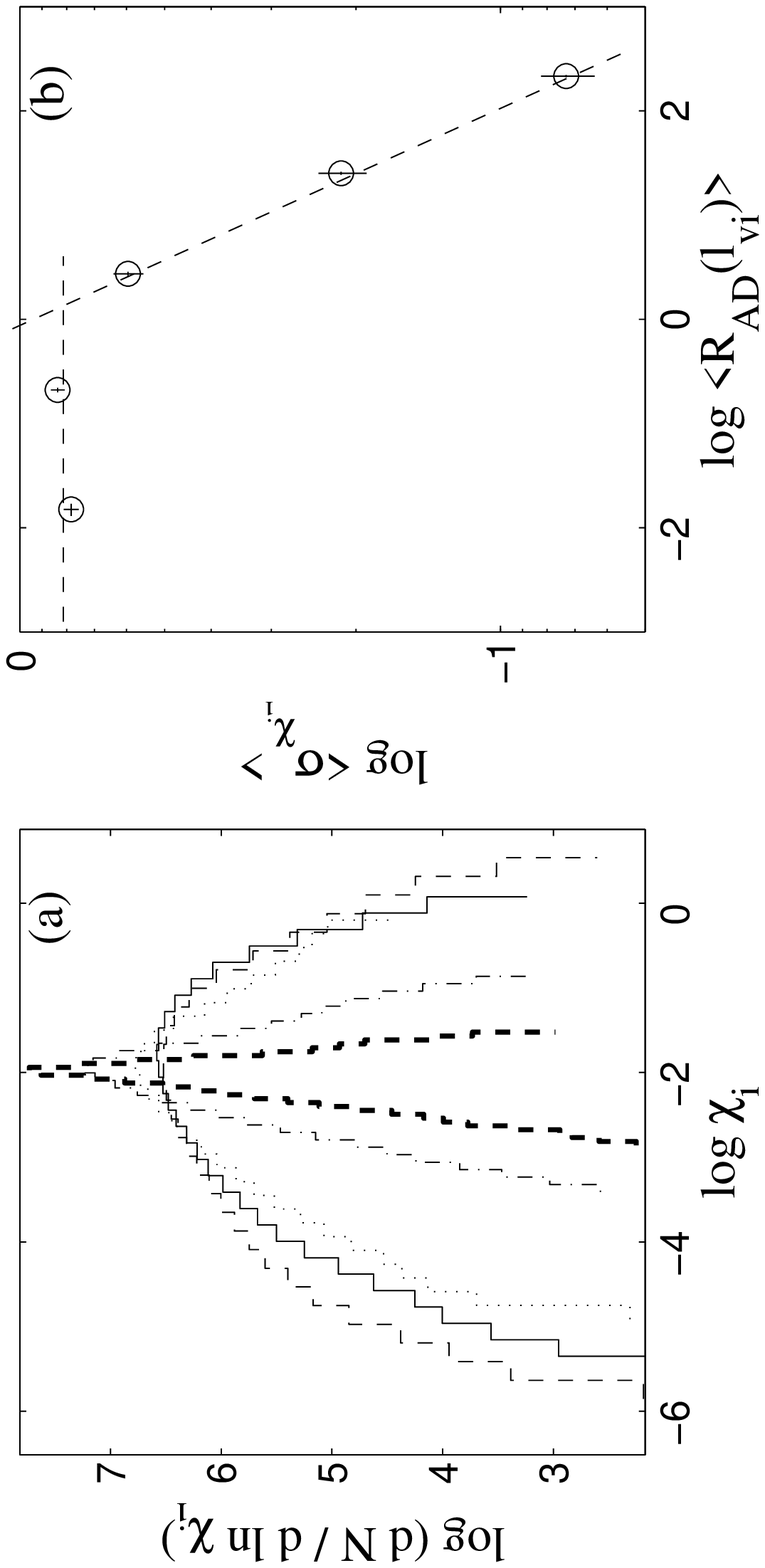}
\caption{$\rad$ effects: (a) The PDFs of $\chi_i$ for models m3c2r-1 (solid), m3c2 (dash), m3c2r1 (dotted), m3c2r2 (dot-dash), and m3c2r3 (thick dash) at t = 3$\tf$ as a function of $\meanradi$.  (b) The time-averaged dispersions of the $\chi_i$ distributions for the five models versus $\meanradi$ over 2 $\tf$.  The $\chi_i$ and $\meanradi$ show a power law relation when the $\meanradi >$ 1.  When $\meanradi <$ 1, the dispersion of $\chi_i$ approaches a constant.
\label{fig11}}
\end{figure}

The dispersion in the ionization changes systematically with $\rad$.  Figure \ref{fig11}a shows the distribution of $\chi_i$ for five models, from m3cr-1 to m3c2r3, at $t = 3\tf$.  We can see that$\chi_i$ has a larger dispersion for smaller values of $\rad$.  This is to be expected, because in the ideal MHD case ($\rad = \infty$), $\chi_i$ will maintain a single value for the whole turbulent box, since the ions and neutrals are perfectly coupled.  With smaller values of $\rad$, the ions start to decouple from the neutrals and the dispersion in $\chi_i$ increases.  We plot the dispersion of $\chi_i$ in Figure \ref{fig11}b.  The dispersion decreases as a power of $\rad$ for models m3c2r1, m3c2r2, and m3cr3, which have $\meanradi > 1$, and then becomes approximately constant for the two models with $\meanradi < 1$.  The turning point is roughly located at $\meanradi \sim 1$. 
The dispersion in the ionization is somewhat larger than the dispersion in the neutral
density, even at the smallest values of $\rad$ we have simulated, since it includes the
dispersion in both the neutral density and the ion density.

\section{Discussion and Conclusions}

Magnetic fields are an important ingredient in the interstellar medium and are believed to play an important role in star formation.  On large scales, the magnetic field is frozen to the gas, and ideal MHD is appropriate.
Indeed, to date almost all 3D simulations of MHD turbulence are based on the assumption of ideal MHD.  On smaller scales, ambipolar diffusion (AD) becomes significant, and in a turbulent medium, the AD lengthscale $\lad$ varies substantially due to high contrasts in density, velocity, and magnetic fields.  When the average value of 
$\lad$ is comparable to or larger than the size of the turbulent box---i.e., when
the AD Reynolds number of the box $\radl\la 1$---AD can significantly alter the
properties of the turbulence.  Simulating this effect is computationally challenging, however, since the timestep required in explicit codes is proportional to both the square of the gridsize, $\Delta x^2$, and to the
square root of the ionization mass fraction, $\chio$, both of which are
exceedingly small for accurate modeling of MCs.
Semi-implicit treatments can avoid the problem with $\Delta x^2$ but not the one with the small ionization mass fraction.
To overcome the latter problem, \citet{ois06} adopted an
artificially high value for the ionization mass fraction, $\chio=0.1$, 
and a correspondingly low value for the ion-neutral
coupling coefficient, $\gad$, in carrying out the 3D simulation for a turbulent medium with ambipolar diffusion. 
\citet{li06} independently developed this approximation, which they termed the heavy-ion approximation, and determined the condition for its validity by testing it on several classical MHD problems involving AD.

In this paper, we report the results of our simulations of sub-\alfvenic turbulence with AD using the heavy-ion approximation.  
We assume that the ions are conserved, but in the Appendix we show that our 
results also apply 
approximately to the case of time-dependent ionization for realistic molecular densities.  Our models focus on the case of a thermal Mach number of 3 and a plasma $\beta$ of 0.1,
corresponding to an \alfven Mach number $\calm_A=0.67$.  
By using this relatively low value of the Mach number, we are able to
perform a number of $256^3$ simulations with a duration of $3-5\tf$, where
$\tf=\ell_0/v_{\rm rms}$ is the flow time across the box.  
We find, in agreement with previous workers \citep[e.g.][]{haw98}, that simulations of turbulence with AD have significant fluctuations in most physical quantities, which makes it difficult to
accurately determine the statistical properties of the system. 
We carried out several convergence studies to determine the validity of our simulations:
First, we showed that independent samples of the turbulence can be obtained at 
time intervals $\simeq t_f$ beginning at $t=\tf$, and that a total running time of
$\sim 3\tf$ is sufficient. Second, we showed that the results are converged to
within the uncertainties for a spatial resolution of $256^3$.
Finally, we showed that the heavy-ion approximation with $\chio = 10^{-2}$ can represent a 
$\chio = 10^{-6}$ system as observed in MCs with sufficient accuracy.  This speeds up the calculation by a factor of about 100, but it is nonetheless a factor 10 slower than an ideal MHD simulation. 

High-resolution ideal MHD turbulence simulations show that the velocity power spectrum for super-\alfvenic turbulence will be close to a Burgers spectrum with a power index $n_v(k)\simeq 2$ (Padoan et al. 2007).  If the turbulence is sub-Alfv$\acute{\rm e}$nic, the power spectrum will be highly anisotropic with respect to the direction of the magnetic field.  Theoretical studies on incompressible turbulence  suggest the velocity power spectrum normal to the magnetic field will be a Kolmogorov-like spectrum, with power index of 5/3 \citep[e.g.][]{gol95}.  Numerical simulations indicate that in strong fields the power index is close to 1.5 (i.e., at low values of $\ma$) \citep{mar01,mul03}, and that the 5/3 index is realized only for relatively weak mean fields \citep{bol05}.

We have studied how the power spectra change as a function of the importance of
ambipolar diffusion, which is measured by
the AD Reynolds number $\rad$. We use two different values of the AD Reynolds number:
$\radl$ is defined in terms of the box size and the rms velocity in the box, whereas
$\meanradi$ is the volume average of the AD Reynolds number defined in terms
of local parameters, with a length equal to the scale over which the ion velocity varies.
$\meanradi$ is the quantity that enters the criterion to ensure the validity of the heavy-ion approximation (eq. \ref{eq:valid}).  It varies with time during a simulation, but is constant to within a factor of 2 in
the simulations reported here; once the system reaches  equilibrium  at  $t\simeq\tf$, the change in $\meanrad$ is $\la10\%$.
For the five models with $\meanradi$ from 0.015 to 215.83 [initial $\radl$ from 0.12 to 1200] that we have computed, we see a progressive transition of system properties from a model with a strong AD effect to a near ideal MHD model; we also computed ideal MHD models for comparison.
All the power law indexes we computed increase in going from the ideal MHD case to the
case of strongest AD, and most of them appear to approach a constant at small $\rad$.  

For the ideal MHD case, we confirm that the power spectrum of the ion velocity normal
to the field is consistent with the Iroshnikov-Kraichnan spectrum  ($n_{vi,r}=1.5$), as found in previous
studies of turbulence in strong fields. This index increases with the importance of AD;
it is consistent with the Kolmogorov value in four of our AD models, but is larger for
the case in which the AD is strongest.
The perpendicular magnetic field power is usually larger than that of the parallel magnetic field power, and as a result the power-law index for the total field, $n_B(k)$, is about the same as that
for the perpendicular components of the field,
$n_{B,r}(k)$. We find that these power-law indexes are about 1.2 in the case of ideal
MHD and rise to about 1.5 for the case in which AD is strongest.
This does not agree with the conclusion of \citet{ois06}, who found no difference in the magnetic power spectra of an ideal MHD model and models with strong AD.  We note that the change in the 
index of the power spectra we find between ideal and AD-dominated MHD is small compared to the difference they reported between ideal MHD and MHD dominated by ohmic diffusion.

By comparing the volume-averaged value of $\ln\rho$, the mass-averaged value, the
dispersion in values, and the median, all of which have equal magnitude for a log-normal PDF,
we concluded that the density PDF is indeed log-normal for all the cases we considered. 
The dispersion increases systematically
as AD increases in importance, and is consistent with the \citet{pad02} result for
the strong AD cases.

An important result from our sub-\alfvenic turbulence simulations with AD is that the neutral gas in systems with small $\beta$ (strong magnetic field) and strong AD (small $\rad$) behaves like that in systems with large $\beta$  (weak magnetic field) and no AD.  In particular, the neutral-velocity power spectrum in a strongly magnetized medium with strong AD is approximately consistent with a Burgers spectrum [$n_v(k)\simeq 2$].  It is thus not possible to infer the strength
of the magnetic field from observations of the power spectrum unless it is known that
the observations are on a sufficiently large scale that AD is not important.

\acknowledgments
We would like to thank the referee for his or her comments on this paper, particularly the suggestion to consider the effect of time-dependent ionization. We thank T. Mouschovias for emphasizing that the additional terms introduced by time dependent ionization are small in comparison with the AD drag terms.
Support for this research was provided by NASA through NASA ATP grant NNG06-GH96G (RIK, CFM, and PSL), under the auspices of the US Department of Energy by Lawrence Livermore National Laboratory under contact DE-AC52-07NA27344 (RIK), and by the NSF through grants AST-0606831 (CFM and RIK) and PHY05-51164 (CFM).  RTF acknowledges support from the DOE ASC/Alliance Center for Astrophysical Thermonuclear Flashes at the University of Chicago, Contract \#B523820.  This research was also supported by grants of high performance computing resources from the San Diego Supercomputer Center and the National Center of Supercomputing Application through grant TG-MCA00N020.

\appendix
\section{ION CONSERVATION VERSUS TIME-DEPENDENT IONIZATION}

The calculations we have discussed in this paper are based on the assumption that
the number of ions is conserved. In fact, as the density changes, ionization and
recombination will change the number of ions. The ionization timescale is
\beq
t_{\rm ion}=\frac{x_e}{\zeta_{\rm CR}}\simeq 100 \left(\frac{x_e}{10^{-7}} \right)\left(\frac{3\times 10^{-17}}{\zeta_{\rm CR}}\right)~~~{\rm yr},
\label{eq:tion}
\eeq
where $x_e\equiv n_e/\nh$ is the ionization number fraction and
$\zeta_{\rm CR}$ is the ionization rate per H atom, which is inferred to be
about $(2.5-5)\times 10^{-17}$~s$^{-1}$ in dense clouds \citep{dal06}. The recombination
time scale is $t_{\rm rec}=1/\alpha n_e$, where $\alpha$ is the relevant recombination coefficient.
In equilibrium these two time scales are equal,
which implies that the equilibrium ionization is
\beq
x_{e,\,\rm eq}=\left(\frac{\zeta_{\rm CR}}{\alpha \nh}\right)^{1/2}\simeq 10^{-7}n_{\rm H,\, 3}^{-1/2},
\eeq
where the numerical evaluation is for $\zeta_{\rm CR}=3\times 10^{-17}$ and
$\alpha = 2.5\times 10^{-6}$~cm$^3$~s\e\ \citep{mck07}.
[This estimate of the ionization is based on the assumption that HCO$^+$ dominates
the ionization; if small PAHs dominate, then the effective recombination rate is about
10 times smaller (\citealp{wak08}) and the equilibrium ionization is several times larger.]
Furthermore, one can show that 
if the ionization is close to equilibrium, then the
the e-folding time for the ionization to approach equilibrium is half as large
as the ionization and recombination times:
\beq
t_{\rm ion,\, eq}=t_{\rm rec,\, eq}=\frac{1}{(\alpha\nh\zeta_{\rm CR})^{1/2}}=2\teq .
\eeq
Note that the ionization time scale is generally orders of magnitude less than the chemical
equilibration time scale, which can be $\sim 10^5$~yr
 \citep[e.g.][]{pad04,wak08}.  In a molecular cloud with low ionization,
the ionization time scale is short compared to the typical dynamical time scale, 
\beq
t_{\rm dyn}=\frac{R}{\sigma}\simeq \frac{R}{0.72R_{\rm pc}^{1/2}~\mbox{km s}^{-1}}
=1.36\times 10^6 R_{\rm pc}^{1/2}~~~\mbox{yr},
\eeq
where $R_{\rm pc}\equiv R/(1~$pc) and where
we have assumed that the velocity dispersion obeys the standard linewidth-size
relation \citep{mck07}. Gas in molecular clouds is therefore expected to be close
to ionization equilibrium except in regions 
where the dynamical time scale is short, as in shocks. 

One can include ionization and recombination as source terms in the mass and momentum equations (\ref{eq:contn})---(\ref{eq:imotion}) as:
\begin{eqnarray}
\frac{\partial \rho_n}{\partial t} &=& - \vecnabla \cdot (\rho_n \vecv_n) - S_1,
\label{eq:contn_ieq}\\
\frac{\partial \rho_i}{\partial t} &=& - \vecnabla \cdot (\rho_i \vecv_i) + S_1,\\
\frac{\partial \rho_n\vecv_n}{\partial t} &=& - \vecnabla\cdot(\rho_n\vecv_n\vecv_n)
 - \grad P_n - \gad \rho_i \rho_n (\vecv_n - \vecv_i) - S_2,
\label{eq:nmotion_ieq}\\
\frac{\partial \rho_i\vecv_i}{\partial t} &=& - \vecnabla\cdot(\rho_i\vecv_i\vecv_i) 
- \grad P_i - \gad \rho_i \rho_n (\vecv_i - \vecv_n) 
+\frac{1}{4 \pi} \left( \curl \vecB \right) \vectimes \vecB + S_2.
\label{eq:imotion_ieq}
\end{eqnarray}
Here the source terms $S_1$ and $S_2$ are
\begin{eqnarray}
S_1 &=& \left( \zeta_{\rm CR} n_H - \alpha n_i^2 \right) m_i \, ,\\
S_2 &=& \zeta_{\rm CR} n_H m_i \left(\vecv_n - \frac{\alpha n_e^2}
{\zeta_{\rm CR} n_H} \vecv_i \right)\, ,
\label{eq:stwo}
\end{eqnarray}
where $m_i$ is the ion mass and we have assumed charge neutrality, $n_i = n_e$.  Since the ionization generally low,  much of the ionization of H$_2$ will be transferred to heavy molecules such as HCO$^+$.  Therefore, we use $m_i$ also in the ionization component of the source terms $S_1$ and $S_2$;
this differs from the treatment in \citet{bra95}.
In equilibrium, $\zeta_{\rm CR} n_H = \alpha n_{e,\, \rm eq}^2$, 
so the coefficient of $\vecv_i$ in equation (\ref{eq:stwo}) is simply 
$(n_e/n_{e,\,\rm eq})^2$. As a result, the second term in the momentum source term dominates
when the gas is overionized.

Let $\crion$ be the ratio of the momentum source term $S_2$ to the AD drag term.
Using equation (\ref{eq:tion}), we find that in equilibrium this ratio is
\beq
\crione = \frac{\zeta n_H m_i}{\gad \rho_i \rho_n} = \frac{1}{2\gad \rho_n t_{\rm eq}}.
\label{eq:ratio}
\eeq
We see that the ionization/recombination source term is important only when the density is very low and/or the ionization timescale is very small. However, in MCs these conditions are generally not satisfied, so that $\crion$ is very small and the ionization/recombination source terms can be ignored.  For example, the typical density and ionization in MCs are $\nht\equiv
\nh/(10^3~\mbox{cm\eee})\ga 1$ and $x_e \sim 10^{-7}\nht^{-1/2}$ \citep{mck07}, 
so that the ionization timescale is
$t_{\rm eq} \sim 60\nht^{-1/2}$~yr and $\crione\sim 1.3 \times 10^{-3}\nht^{-1/2}$.

To verify that the momentum source terms are indeed negligible in realistic cases, we modified ZEUS-MPAD to include the source terms and performed a model simulation based on the initial conditions of model m3c2 (which has $\crione=0$) but with 
$\crione= 1.7\times 10^{-3}$. 
This corresponds to a density $\nh=550$~cm\eee; since
this is smaller than the typical density in molecular gas, this represents
an approximate upper bound on the effect of time-dependent ionization.  The mean value of $\crione$ in the model is about a factor of 1.6
times the initial value using equation (\ref{eq:ratio}): 
$\crione\propto 1/\rho_n^{1/2}$, and for a log-normal distribution
one can show that the mean of 
$(\bar\rho_n/\rho_n)^{1/2}$ is $\exp(3\sigma^2/8)$,
where $\sigma^2$ is the dispersion of the log normal. 
\citet{pad02} estimate $\sigma^2\simeq  \ln[1 + (\calm/2)^2]$, which is 1.18 for our Mach 3 models; hence, $\langle\crione\rangle\simeq 2.8\times 10^{-3}$.
However, non-equilibrium effects are very important: 
For those cells that are overionized, 
$\crion$ will be larger by a factor of order $\alpha n_e^2/\zeta_{\rm CR} n_H = (n_e/n_{e,{\rm eq}})^2$.  The average value of
$(n_e/n_{e,\,\rm eq})^2$ is about 16, and as a result the average value of $\crion$ is
0.02, significantly larger than the equilibrium value.
Although the mean value of $\crion$ is small, time-dependent ionization has
a detectable effect on the spectra of the turbulence, being about $1 \;\sigma$ steeper
than those for the case of ion conservation. For realistic molecular densities, 
$\crion$ will be smaller, so the spectra for the time-dependent case will be closer 
to those for the conservation case.
We conclude that MC models with time-dependent ionization are generally well approximated by models using the assumption of ion conservation.

\clearpage
\clearpage
\clearpage
\clearpage
\clearpage
\clearpage
\clearpage
\clearpage
\clearpage
\clearpage
\clearpage
\clearpage
\begin{table}
\begin{center}
\caption{Model Parameters and Statistical Results\label{tbl-1}}
\begin{tabular}{lcccccccc}
\\
\tableline\tableline
Model$^*$ & $\calm_{\rm rms}$ & $\chi_i$ & $\gad$ & Time & $\rad(\ell_0)$ & $\avg{\rad(\ell_{v_n})}_V^{\dagger\dagger}$ & $\avg{\rad(\ell_{v_i})}_V^{\dagger\dagger}$ & $\calm_{\rm Ai}^2/\avg{\rad(\ell_{v_i})}_V^{\dagger\dagger}$ \\
 & & & &($\tf$) & & & & ($\times 10^{-3}$)\\
\tableline
m3c1   &3  &10$^{-1}$ &4        &2    &1.2      &0.3916   &0.2184   &167.5\\
m3c2   &3  &10$^{-2}$ &40       &3    &1.2      &0.3206   &0.2197   &16.01\\
m3c3   &3  &10$^{-3}$ &400      &2    &1.2      &0.3392   &0.1908   &1.849\\
m3c4   &3  &10$^{-4}$ &4000     &2    &1.2      &0.3363   &0.1894   &0.213\\
m3c2a$^{**}$&3  &10$^{-2}$ &40  &3    &1.2      &0.3374   &0.2183   &15.17\\
m3c2h$^{***}$&3 &10$^{-2}$ &40  &3    &1.2      &0.2709   &0.1750   &20.17\\
m3c2r-1&3  &10$^{-2}$ &4        &3    &0.12     &0.0298   &0.0145   &148.1\\
m3c2r1 &3  &10$^{-2}$ &400      &3    &12       &2.9266   &2.5480   &1.767\\
m3c2r2 &3  &10$^{-2}$ &4000     &3    &120      &25.455   &25.090   &0.179\\
m3c2r3 &3  &10$^{-2}$ &40000    &3    &1200     &227.47   &225.59   &0.020\\
m10c1  &10 &10$^{-1}$ &4        &1.25 &4        &1.0720   &0.4326   &909.6\\
m10c2  &10 &10$^{-3}$ &40       &1.25 &4        &1.1596   &0.4172   &90.24\\
m10c3  &10 &10$^{-4}$ &400      &1.25 &4        &1.2653   &0.4249   &8.860\\
m3i$^{\dagger}$ &3  &$\infty$  &$\infty$ &3    &$\infty$ &$\infty$ &$\infty$ &- \\
m3ih$^{\dagger***}$&3  &$\infty$  &$\infty$ &3    &$\infty$ &$\infty$ &$\infty$ &- \\
m10i$^{\dagger}$&3  &$\infty$  &$\infty$ &3    &$\infty$ &$\infty$ &$\infty$ &- \\
\tableline
\end{tabular}
\end{center}
$^*$ Models are labeled as ``mxcy," where $x$ is the thermal Mach number and $y=|\log\chio|$.
Models labeled ``mxcyrn" have $\radl=1.2\times 10^n$.

$^{**}$ Driving applied only to the neutrals.

$^{***}$ High resolution model  ($512^3$).

$^{\dagger}$ Ideal MHD models.

$^{\dagger\dagger}$ Root mean squared (rms) values.
\end{table}
\clearpage

\begin{deluxetable}{lccccccccc}
\rotate
\tablecaption{Spectral indexes of Velocity and Magnetic Field Power Spectra for Models in Convergence Studies \label{tbl-2}}
\tablewidth{0pt}
\tablehead{
\colhead{Model} & \colhead{$n_{\rm vi,r}(k)$} & \colhead{$n_{\rm vi,z}(k)$} & \colhead{$n_{\rm vi}(k)$} & \colhead{$n_{\rm vn,r}(k)$} & \colhead{$n_{\rm vn,z}(k)$} & \colhead{$n_{\rm vn}(k)$} & \colhead{$n_{\rm B,r}(k)$} & \colhead{$n_{\rm B,z}(k)$} & \colhead{$n_{\rm B}(k)$}
}
\startdata
m3c1  &1.87$\pm$0.19 &1.87$\pm$0.09 &1.87$\pm$0.10 &1.99$\pm$0.06 &1.99$\pm$0.10 &1.99$\pm$0.06 &1.41$\pm$0.16 &2.28$\pm$0.19 &1.55$\pm$0.16\\
m3c2  &1.79$\pm$0.14 &1.92$\pm$0.07 &1.88$\pm$0.08 &1.92$\pm$0.07 &1.90$\pm$0.07 &1.91$\pm$0.06 &1.44$\pm$0.12 &2.12$\pm$0.13 &1.55$\pm$0.12\\
m3c3  &1.75$\pm$0.18 &1.91$\pm$0.09 &1.86$\pm$0.07 &1.94$\pm$0.04 &1.90$\pm$0.09 &1.93$\pm$0.04 &1.40$\pm$0.12 &2.02$\pm$0.18 &1.50$\pm$0.13\\
m3c4  &1.79$\pm$0.16 &1.90$\pm$0.06 &1.82$\pm$0.11 &1.91$\pm$0.07 &1.90$\pm$0.07 &1.91$\pm$0.05 &1.42$\pm$0.12 &2.08$\pm$0.16 &1.53$\pm$0.13\\
m3c2h &1.64$\pm$0.06 &1.98$\pm$0.03 &1.86$\pm$0.03 &1.92$\pm$0.03 &1.95$\pm$0.03 &1.94$\pm$0.03 &1.48$\pm$0.05 &2.14$\pm$0.07 &1.56$\pm$0.05\\
m3i   &1.46$\pm$0.10 &1.31$\pm$0.10 &1.41$\pm$0.08 & - & - & - &1.17$\pm$0.09 &1.72$\pm$0.11 &1.25$\pm$0.09\\
m3ih  &1.48$\pm$0.06 &1.38$\pm$0.05 &1.45$\pm$0.05 & - & - & - &1.14$\pm$0.07 &1.61$\pm$0.08 &1.23$\pm$0.07\\
m3c2a &1.83$\pm$0.14 &1.93$\pm$0.07 &1.90$\pm$0.06 &1.93$\pm$0.05 &1.90$\pm$0.07 &1.92$\pm$0.04 &1.40$\pm$0.10 &2.03$\pm$0.16 &1.50$\pm$0.10\\
m10c1 &1.50 &1.73 &1.57 &2.00 &1.79 &1.94 &0.99 &1.29 &1.05\\
m10c2 &1.05 &1.86 &1.34 &2.00 &1.93 &1.98 &1.11 &1.63 &1.21\\
m10c3 &1.05 &1.73 &1.34 &2.04 &1.84 &1.98 &1.15 &1.19 &1.16\\
m10i  &1.11 &1.53 &1.27 & - & - & - &1.38 &1.18 &1.34\\
\enddata
\end{deluxetable}
\clearpage

\begin{table}
\begin{center}
\caption{Spectral indexes of Velocity and Magnetic Field Power Spectra for $\calm$ = 3 Models with Varying $\rad$ \label{tbl-3}}
\begin{tabular}{lcccccccc}
\\
\tableline\tableline
                      & m3c2r-1        & m3c2          & m3c2r1        & m3c2r2         & m3c2r3          & m3i\\
\tableline
$\rad(\ell_0)$        & 0.12           & 1.2           & 12.0          &
120            & 1200            & $\infty$\\
$\avg{\rad(\ell_{v_i})}_V$ & 0.015$\pm$0.002& 0.21$\pm$0.01& 2.74$\pm$0.13 & 25.25$\pm$0.79 & 215.83$\pm$5.00& $\infty$\\
$n_{\rm vi,r}(k)$     & 1.89$\pm$0.15  & 1.79$\pm$0.14 & 1.78$\pm$0.13 & 1.72$\pm$0.12  & 1.67$\pm$0.10   & 1.46$\pm$0.10\\
$n_{\rm vi,z}(k)$     & 2.07$\pm$0.06  & 1.92$\pm$0.07 & 1.80$\pm$0.14 & 1.32$\pm$0.12  & 1.27$\pm$0.16 & 1.31$\pm$0.10\\
$n_{\rm vi}(k)$       & 1.92$\pm$0.08  & 1.88$\pm$0.08 & 1.82$\pm$0.12 & 1.56$\pm$0.08  & 1.50$\pm$0.09   & 1.41$\pm$0.08\\
$n_{\rm vn,r}(k)$     & 2.17$\pm$0.05  & 1.92$\pm$0.07 & 1.94$\pm$0.09 & 1.85$\pm$0.12  & 1.67$\pm$0.10   & - \\
$n_{\rm vn,z}(k)$     & 2.07$\pm$0.06  & 1.90$\pm$0.07 & 1.80$\pm$0.07 & 1.32$\pm$0.12  & 1.27$\pm$0.16   & - \\
$n_{\rm vn}(k)$       & 2.14$\pm$0.04  & 1.91$\pm$0.06 & 1.87$\pm$0.09 & 1.57$\pm$0.15  & 1.50$\pm$0.09   & - \\
$n_{\rm B,r}(k)$      & 1.45$\pm$0.10  & 1.44$\pm$0.12 & 1.38$\pm$0.15 & 1.27$\pm$0.13  & 1.12$\pm$0.08   & 1.17$\pm$0.09\\
$n_{\rm B,z}(k)$      & 2.03$\pm$0.12  & 2.12$\pm$0.13 & 2.08$\pm$0.12 & 2.03$\pm$0.13  & 1.82$\pm$0.08   & 1.72$\pm$0.11\\
$n_{\rm B}(k)$        & 1.53$\pm$0.09  & 1.55$\pm$0.12 & 1.50$\pm$0.14 & 1.38$\pm$0.12  & 1.22$\pm$0.08   & 1.25$\pm$0.09\\
$n_{\rm vi}(k_r)$     & 1.95$\pm$0.09  & 1.99$\pm$0.06 & 1.96$\pm$0.13 & 1.64$\pm$0.08  & 1.54$\pm$0.09   & 1.60$\pm$0.07\\
$n_{\rm vn}(k_r)$     & 2.11$\pm$0.04  & 2.08$\pm$0.06 & 1.83$\pm$0.09 & 1.66$\pm$0.08  & 1.55$\pm$0.09   & - \\
$n_{\rm B}(k_r)$      & 1.50$\pm$0.10  & 1.53$\pm$0.11 & 1.36$\pm$0.15 & 1.29$\pm$0.13  & 1.29$\pm$0.08   & 1.33$\pm$0.09\\
\tableline
\end{tabular}
\end{center}
\end{table}
\clearpage
\begin{table}
\begin{center}
\caption{Statistical Parameters of the Density PDF \label{tbl-4}}
\begin{tabular}{cccccc}
\\
\tableline\tableline
Model  & $\avg{\rad(\ell_{v_i})}_V$ & $-\avg{x}_V$  & $\avg{x}_M$   & $-\tilde{x}$      & $\frac{1}{2}\sigma_x^2$\\
\tableline
m3c2r-1& 0.015$\pm$0.002 & 0.58$\pm$0.03 & 0.55$\pm$0.02 & 0.56$\pm$0.04 & 0.60$\pm$0.03\\
m3c2   & 0.21$\pm$0.01   & 0.65$\pm$0.04 & 0.60$\pm$0.04 & 0.62$\pm$0.05 & 0.70$\pm$0.05\\
m3c2r1 & 2.74$\pm$0.13   & 0.59$\pm$0.07 & 0.51$\pm$0.05 & 0.49$\pm$0.07 & 0.70$\pm$0.12\\
m3c2r2 & 25.25$\pm$0.79  & 0.32$\pm$0.03 & 0.32$\pm$0.03 & 0.33$\pm$0.03 & 0.32$\pm$0.02\\
m3c2r3 & 215.83$\pm$5.00 & 0.39$\pm$0.03 & 0.37$\pm$0.03 & 0.38$\pm$0.05 & 0.40$\pm$0.02\\
m3c2h  & 0.18$\pm$0.01   & 0.60$\pm$0.03 & 0.56$\pm$0.02 & 0.58$\pm$0.02 & 0.64$\pm$0.05\\
m3i    & $\infty$        & 0.43$\pm$0.04 & 0.40$\pm$0.04 & 0.40$\pm$0.07 & 0.46$\pm$0.04\\
m3ih   & $\infty$        & 0.40$\pm$0.05 & 0.38$\pm$0.06 & 0.39$\pm$0.09 & 0.42$\pm$0.05\\
\tableline
\end{tabular}
\end{center}
\end{table}
\clearpage

\end{document}